\documentclass[12pt, draftclsnofoot, onecolumn]{IEEEtran}
\usepackage{epsfig,graphicx,subfigure,psfrag,amsmath,cases}
\usepackage{latexsym,amssymb,amsmath,epsfig,subfigure,algorithm,mathtools}
\usepackage{algorithmic}
\usepackage{color}
\usepackage{url}
\usepackage{scrtime}
\usepackage{array}

\author{\authorblockN{Derrick Wing Kwan Ng, Yongpeng Wu, and Robert Schober\thanks{Derrick Wing Kwan Ng,  Yongpeng Wu, and Robert Schober are with the Institute for Digital Communications (IDC),
Friedrich-Alexander-University Erlangen-N\"urnberg (FAU), Germany (email:\{kwan, yongpeng.wu, schober\}@lnt.de). Derrick Wing Kwan Ng and Robert Schober are also with the University of British Columbia, Vancouver, Canada.}}}

\title{\vspace*{-10mm}Power Efficient Resource Allocation for Full-Duplex Radio Distributed Antenna Networks}

\date{\thistime,\,\today}

\newtheorem{Thm}{Theorem}

\newtheorem{Prop}{Proposition}
\newtheorem{Remark}{Remark}
 \newcommand{\qed}{\hfill \ensuremath{\blacksquare}}
 
 \newcommand{\textoverline}[1]{$\widetilde{\mbox{#1}}$}
\DeclareMathOperator{\Tr}{\mathrm{Tr}}
\DeclareMathOperator{\zero}{\mathbf{0}}
\DeclareMathOperator{\Rank}{\mathrm{Rank}}

\DeclareMathOperator{\diag}{\mathrm{diag}}

\DeclareMathOperator{\maxo}{maximize}
\DeclareMathOperator{\mino}{minimize}
\DeclareMathOperator{\bigo}{\cal O}
\newcommand{\abs}[1]{\lvert#1\rvert}
\newcommand{\norm}[1]{\lVert#1\rVert}

\newcolumntype{L}{>{\arraybackslash\raggedright}m{11cm}}
\makeatletter

\newcommand{\Rmnum}[1]{\expandafter\@slowromancap\romannumeral #1@}
\makeatother
\begin{document}

\maketitle
\vspace*{-20mm}
\begin{abstract}
In this paper, we study the  resource allocation algorithm design  for distributed antenna multiuser networks  with full-duplex (FD) radio base stations (BSs) which enable simultaneous uplink and downlink  communications.  The considered resource allocation algorithm design is formulated as an optimization problem taking into account the antenna circuit power
consumption of the BSs and the quality of service (QoS) requirements
of  both uplink and downlink users. We minimize the total network power consumption by jointly optimizing the downlink beamformer, the uplink transmit power, and the antenna selection.   To overcome the intractability of the resulting problem, we reformulate it as an optimization problem with decoupled binary selection variables and  non-convex constraints.  The reformulated problem facilitates the design of an
iterative resource allocation algorithm which obtains an optimal solution based on the generalized Bender's decomposition (GBD) and serves as a benchmark scheme.  Furthermore,  to strike a balance between computational complexity and system performance, a suboptimal algorithm with polynomial time complexity is proposed.  Simulation results illustrate that the proposed GBD based iterative algorithm converges to the global optimal solution and the suboptimal algorithm achieves a close-to-optimal performance. Our results also demonstrate
 the trade-off between power efficiency and the number of active transmit antennas when
 the circuit power consumption is taken into account. In particular,  activating an exceedingly large number of antennas may not be a power efficient solution for reducing
 the total system power consumption. In addition, our results reveal that FD systems   facilitate  significant power savings compared to traditional half-duplex systems, despite the non-negligible self-interference.

\end{abstract}
\vspace*{-4mm}
\vspace*{-5mm}
\section{Introduction} \label{sect1}
The next generation wireless communication systems are required to support ubiquitous and high data rate communication applications with  guaranteed quality of service (QoS). These requirements translate into a tremendous  demand for  bandwidth and energy consumption. Multiple-input multiple-output (MIMO) is a viable solution for addressing these issues as it provides extra degrees of freedom in the spatial domain which facilitates a trade-off between multiplexing
gain and diversity gain.  Hence, a large amount of work has been  devoted to  MIMO communication over the
past decades \cite{book:david_wirelss_com,JR:limited_backhaul_vincent}.  However, the
modest computational capabilities of  mobile devices  limit the MIMO gains that can be
achieved in practice.  An attractive
alternative for realizing the performance gains offered by multiple antennas is multiuser MIMO,
 where a multiple-antenna transmitter serves multiple single-antenna receivers simultaneously  \cite{JR:large_number_antennas,JR:TWC_large_antennas}. In fact, the combination of multiuser MIMO and distributed antennas is widely recognized as a promising technology for mitigating interference and  extending service coverage \cite{JR:DAS_MU-MIMO}--\nocite{JR:DAS_MU-MIMO2}\cite{JR:comp}.  Specifically, distributed antennas   introduce additional  capabilities for combating both  path loss and shadowing by  shortening  the distances between the transmitters and the receivers. Nevertheless, if the number of antennas is very large,  the circuit power consumption of  distributed antenna networks  becomes non-negligible  compared to the power consumed for transmission. However, this problem has not been considered in most of the existing literature  \cite{JR:DAS_MU-MIMO}--\nocite{JR:DAS_MU-MIMO2}\cite{JR:comp} on power efficient communication network design. Furthermore,  even with these powerful MIMO techniques, spectrum scarcity is still a major obstacle in providing high speed uplink and downlink communications.

Traditional communication systems  are designed for half-duplex (HD) transmission  since this mode of operation facilitates  low-complexity transceiver design.
In particular, uplink and downlink communication are statically separated in either time or frequency, e.g. via time division duplex  or frequency division duplex,  which leads to a
loss in spectral efficiency.  Even though different approaches have been proposed for improving
the spectral efficiency of HD systems, e.g. dynamic uplink-dowlink scheduling/allocation in time division duplex communication systems \cite{CN:dynamic_TDD,CN:dynamic_TDD2}, the fundamental spectral efficiency loss induced by the HD constraint remains unsolved.  On the contrary, full duplex (FD) transmission  allows  downlink and uplink transmission to occur simultaneously at the same frequency. In fact, FD radio has the potential to double the spectral efficiency of conventional HD communication systems. However, in practice, the downlink transmission in FD systems creates  self-interference to the uplink receive antennas which can be exceedingly large compared to the received power of the useful information signals. In fact, the huge difference in the power levels of the two signals  saturates  the dynamic range of the analog-to-digital converter (ADC) essentially preventing FD communication. Fortunately, several recent breakthroughs in hardware (/signal processing algorithm) design for suppressing self-interference have been reported and FD radio prototypes  have been   successfully  presented \cite{CN:Full_duplex_radio}--\nocite{CN:Full_duplex_radio1,CN:standford_FD,CN:Full_duplex_radio2}\cite{JR:Full_duplex_radio1}. As a result, FD radio has regained
the attention of both industry \cite{patent:FD1}\nocite{3Gpp:Full_duplex1,3Gpp:Full_duplex2}--\cite{Project:Full_duplex3} and academia \cite{JR:Full_duplex_radio2}--\nocite{JR:Full_duplex_radio3,JR:Kwan_FD,JR:FD_DC_program2}\cite{JR:FD_large_antennas}.  In \cite{JR:Full_duplex_radio2}, the authors studied techniques for self-interference suppression and  cancellation for FD multiple-antenna relays. In \cite{JR:Full_duplex_radio3}, the outage probability of MIMO FD single-user relaying systems was investigated. In \cite{JR:Kwan_FD}, a resource allocation algorithm was proposed for maximization of the achievable end-to-end system  data rate of multicarrier multiuser MIMO FD relaying systems.  In \cite{JR:FD_DC_program2}, a suboptimal beamformer design was considered to improve the spectral efficiency of a FD radio base station enabling simultaneous uplink and downlink communication. In \cite{JR:FD_large_antennas}, the concept of FD communication was extended to the case of massive MIMO where a  FD radio relay is equipped with a large number of antennas for suppressing the self-interference and for enhancing the system throughput. However, the benefits of multiple-antenna FD radio do not come for free.  The rapidly escalating cost caused by the  power consumption of the circuitries of large antenna systems has lead to significant financial implications for service providers, which is often overlooked in the literature
\cite{CN:Full_duplex_radio}-\cite{JR:FD_large_antennas}. In fact, the systems in \cite{CN:Full_duplex_radio}-\cite{JR:FD_large_antennas} are designed to serve peak service demands by activating all  available antennas of the system,  without considering the power consumption in the off-peak periods. However,  the service loads vary across a wireless network in practice, depending on the geographic location of the receivers and the time of day. Thus, we expect that extra power savings can be achieved  by dynamically switching off some of the antennas.  Nevertheless, the optimal number of
active antennas   has not been investigated from a system power efficiency
point of view for FD radio communication, yet. In addition, there may be fewer degrees of freedom for self-interference suppression at each FD radio base station in distributed antenna systems if the total number of antennas in the network is fixed. Thus, it is unclear whether the  distributed antenna architecture leads to power savings for FD radio communication. Furthermore, unlike for the orthogonal transmission adopted in HD systems,  the uplink and downlink transmit powers are  coupled  in FD systems which make the design of efficient resource allocation algorithms particularly challenging.

In this paper, we address the above issues and study the resource allocation algorithm design for multiuser distributed antenna  communication networks. We minimize the total network power consumption while  taking into account the circuit power consumption of the distributed BS antennas and ensuring the QoS of both uplink and downlink users.  In particular, we propose an optimal iterative resource allocation algorithm based on the generalized Bender's decomposition \cite{JR:Vijay_GBD}--\nocite{book:non_linear_and_mixed_integer,JR:generalized_Bender's}\cite{book:non_linear_integer}. Furthermore, we propose a suboptimal resource allocation scheme with polynomial time computational complexity  based on the difference of convex functions (d.c.) programming \cite{JR:DC_programming}  which finds a local optimal solution for the considered optimization problem.

\vspace*{-6mm}
\section{System Model}
\label{sect:system model}\vspace*{-2mm}
\subsection{Notation}
 Matrices and vectors are represented by boldface capital and lower case letters, respectively. $\mathbf{A}^H$,  $\mathbf{A}^T$, $\Tr(\mathbf{A})$, and $\Rank(\mathbf{A})$ represent the  Hermitian transpose, the transpose,   the trace, and the rank of  matrix $\mathbf{A}$, respectively; $\mathbf{A}\succ \mathbf{0}$ and $\mathbf{A}\succeq \mathbf{0}$ indicate that $\mathbf{A}$ is a positive definite and a  positive semidefinite matrix, respectively;  $\mathbf{I}_N$ is the $N\times N$ identity matrix; $\mathbb{C}^{N\times M}$ and $\mathbb{H}^N$  denote the sets of all $N\times M$ matrices and $N\times N$ Hermitian matrices with complex entries, respectively;  $\diag(x_1, \cdots, x_K)$ denotes a diagonal matrix with the diagonal elements given by $\{x_1, \cdots, x_K\}$; $\abs{\cdot}$ denotes the absolute value of a complex scalar; the circularly symmetric complex Gaussian distribution is denoted by ${\cal CN}(\boldsymbol{\mu},\mathbf{C})$ with mean vector $\boldsymbol{\mu}$ and co-variance matrix $\mathbf{C}$; $\sim$ stands for ``distributed as";  ${\cal E}\{\cdot\}$ denotes statistical expectation;  and
 $\nabla_{\mathbf{x}} f$ denotes the gradient of a  function $f$ with respect to vector $\mathbf{x}$.
 \vspace*{-4mm}
\subsection{System Model }
\label{sect:multicell-central-unit}
\begin{figure}[t]
\centering\vspace*{-18mm}
\includegraphics[width=3.6 in]{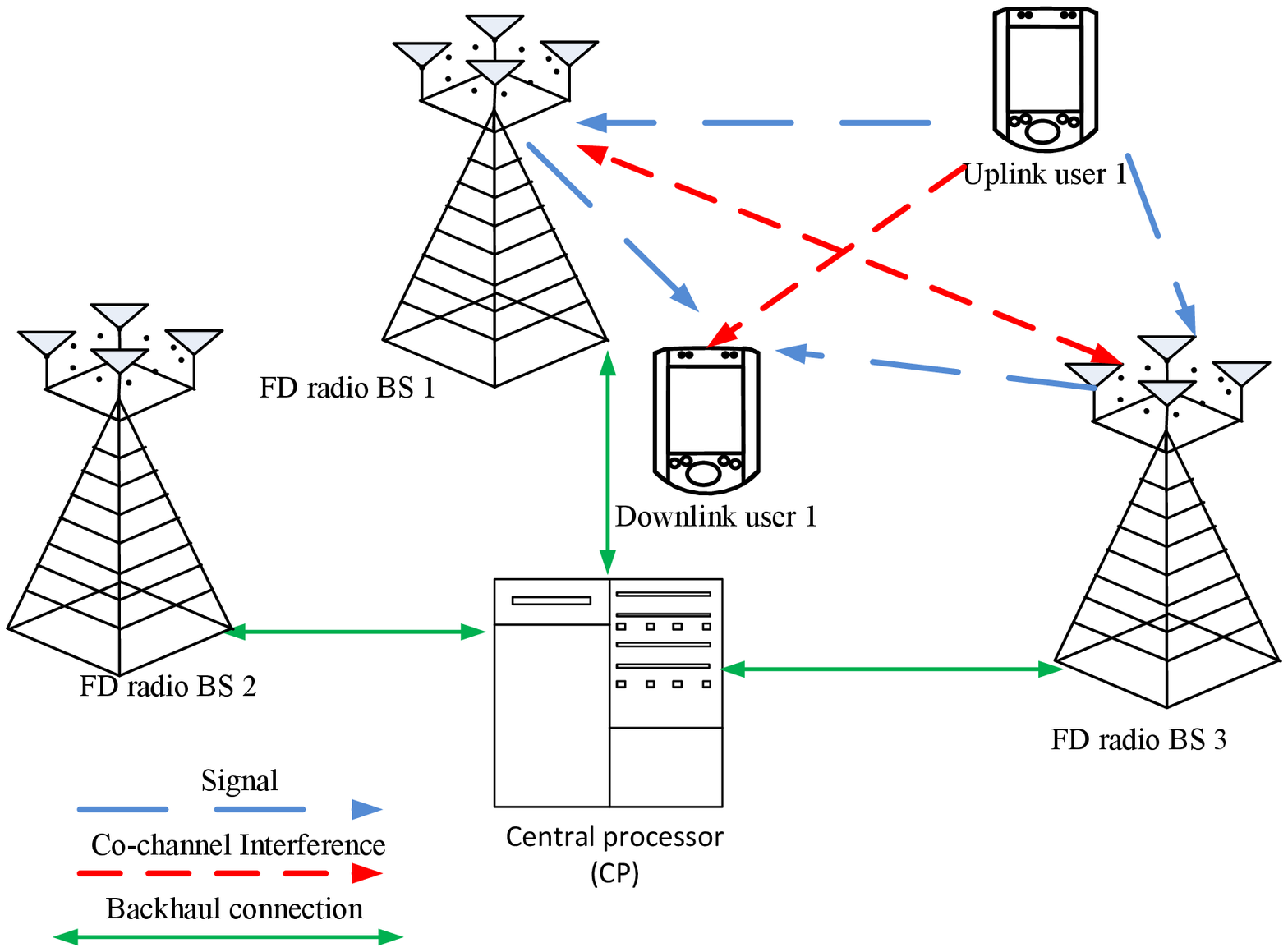}\vspace*{-6mm}
\caption{Multiuser downlink distributed antenna communication system model with $L=3$ full duplex (FD) radio base stations (BSs), $K_{\mathrm{U}}=1$ uplink user, and  $K_{\mathrm{D}}=1$ downlink user. For the depicted case, the antennas equipped at FD radio BS $2$ are switched to idle mode for reducing the total power consumption in the network. }
\label{fig:system_model}\vspace*{-8mm}
\end{figure}

We consider a distributed antenna multiuser communication network. The system consists of a central processor (CP),  $L$ FD radio base stations (BSs), and  $K$ mobile users, cf. Figure \ref{fig:system_model}. Each FD radio BS is equipped with $N_\mathrm{T}>1$  antennas for downlink transmission and uplink reception\footnote{We assume that the  antennas equipped at the FD BSs can transmit and receive simultaneously which has been successfully demonstrated in some FD radio prototypes \cite{CN:standford_FD}.  }. The $K$ users employ single-antenna HD mobile communication devices to ensure low hardware complexity.  In particular, $K_{\mathrm{U}}$ and $K_{\mathrm{D}}$  users are scheduled for simultaneous uplink and downlink transmission, respectively, such that $K_{\mathrm{U}}+K_{\mathrm{D}}=K$. On the other hand, the CP is the core unit of the network. In particular, the FD radios are connected to the CP via backhaul links. In addition, the CP has the full channel state information of the entire network and the data of all downlink users for resource allocation. In this paper, we assume that the CP is a powerful computing unit, e.g. a series of baseband units as in cloud radio access networks (C-RAN), which computes the
resource allocation policy and broadcasts it to all FD radio BSs. Each FD radio BS receives the control signals for resource allocation and the data of the $K_{\mathrm{D}}$ downlink users from the CP via a backhaul link. Furthermore, the FD radio BSs transfer the received uplink signals via backhaul links  to the CP, where  the information is decoded. In this paper, we assume that the backhaul links are implemented with optical fiber and have sufficiently large capacity and low latency to support  real time information exchange between the CP and the FD radio BSs.  For studies on the impact of a limited backhaul capacity on the performance of wireless systems, please refer to \cite{JR:limited_backhaul_vincent,JR:Kwan_limited_backhual}.
\vspace*{-6mm}
\subsection{Channel Model}
A frequency flat fading channel is assumed\footnote{The frequency flat fading channel can be interpreted as one subcarrier of an orthogonal frequency division multiplexing system.  } in this paper.
 The received signals at downlink user $k\in\{1,\ldots,K_{\mathrm{D}}\}$ and the $L$ FD radio BSs are given by \vspace*{-2mm}
\begin{eqnarray}
y_{k}^{\mathrm{DL}}&=&\mathbf{h}_{\mathrm{D}_k}^H\mathbf{x}+\underbrace{\sum_{j=1}^{K_{\mathrm{U}}}\sqrt{P^{\mathrm{U}}_j}g_{j,k}d^{\mathrm{U}}_j}_{
\mathrm{co-channel\, interference}}+ n_{k}\quad \mbox{and}\\ \mathbf{y}^{\mathrm{UL}}&=&\sum_{j=1}^{K_{\mathrm{U}}}P^{\mathrm{U}}_j\mathbf{h}_{\mathrm{U}_j}d^{\mathrm{U}}_j+\underbrace{\mathbf{H}_{\mathrm{SI}}\mathbf{x}}_{\mathrm{self-interference}}+\mathbf{z},\,\,
\end{eqnarray}
respectively, where $\mathbf{x}\in\mathbb{C}^{N_{\mathrm{T}}L\times1}$ denotes the joint transmit signal vector of the $L$ FD radio BSs to the $K_{\mathrm{D}}$ downlink users.  The downlink channel between the $L$ FD radio BSs  and  user $k$ is denoted by $\mathbf{h}_{\mathrm{D}_k}\in\mathbb{C}^{N_{\mathrm{T}}L\times1}$, and we use ${g}_{j,k}\in\mathbb{C}$ to represent the channel  between uplink user $j$  and downlink user $k$.  $d^{\mathrm{U}}_j$ and $P^{\mathrm{U}}_j$ are the transmit data and transmit power sent from uplink user $j$ to the $L$ FD radio BSs, respectively. $\mathbf{h}_{\mathrm{U}_j}\in\mathbb{C}^{N_{\mathrm{T}}L\times1}$ is the uplink channel between uplink user $j$ and the  $L$ FD radio BSs. Due to simultaneous uplink reception and downlink  transmission at the FD radio BSs, self-interference from the downlink impairs the uplink signal reception. In practice, different interference mitigation techniques  such as antenna cancellation, balun cancellation, and circulators \cite{CN:standford_FD,CN:Full_duplex_radio2} have been proposed to alleviate the impairment caused by self-interference. In order to isolate the resource allocation algorithm design from the specific implementation of self-interference mitigation, we model the  residual self-interference after interference cancellation by matrix $\mathbf{H}_{\mathrm{SI}}\in\mathbb{C}^{N_{\mathrm{T}}L\times N_{\mathrm{T}}L}$. Variables  $\mathbf{h}_{\mathrm{D}_k}$, ${g}_{j,k}$, $\mathbf{H}_{\mathrm{SI}}$, and $\mathbf{h}_{\mathrm{U}_j}$ capture the joint effect of path loss and multipath fading. $\mathbf{z}\sim{\cal CN}(\zero,\sigma_{\mathrm{z}}^2\mathbf{I}_{N_{\mathrm{T}}L})$ and $n_{k}\sim{\cal CN}(0,\sigma_{\mathrm{n}_k}^2)$  represent the additive white Gaussian noise (AWGN) at the $L$ FD radio BSs and user $k$, respectively.

In each scheduling time slot, $K_{\mathrm{D}}$ independent signal streams are transmitted simultaneously at the same frequency to the $K_{\mathrm{D}}$ downlink users. Specifically,  a dedicated downlink beamforming weight, ${w}_k^l\in\mathbb{C}$, is allocated to downlink user $k$ at the $l$-th, $l\in\{1,\ldots,N_{\mathrm{T}}L\}$, antenna to facilitate downlink information transmission. For  the sake of presentation, we define a  super-vector $\mathbf{w}_k\in\mathbb{C}^{N_{\mathrm{T}}L\times 1}$  for downlink user $k$ as
\begin{eqnarray}
\mathbf{w}_k=\big[{w}_k^1 \,{w}_k^2\,\ldots\,{w}_k^{N_{\mathrm{T}}L}\big]^T.
\end{eqnarray}
  $\mathbf{w}_k$ represents the joint beamformer used by the $N_{\mathrm{T}}L$ antennas shared by the FD radio BSs for serving downlink user $k$. Then, the information signal to downlink user $k$,  $\mathbf{x}_k$, can be expressed as
\begin{eqnarray}
\mathbf{x}_k=\mathbf{w}_kd^{\mathrm{D}}_k,
\end{eqnarray}
where $d^{\mathrm{D}}_k\in\mathbb{C}$ is the data symbol for downlink user $k$. Without loss of generality, we assume that
 ${\cal E}\{\abs{d_k^{\mathrm{D}}}^2\}={\cal E}\{\abs{d_j^{\mathrm{U}}}^2\}=1,\forall k\in\{1,\ldots,K_{\mathrm{D}}\},j\in\{1,\ldots,K_{\mathrm{U}}\}$.
\subsection{Network Power Consumption Model}
In our system model, we include the circuit power
consumption of the system in the  objective
function in order to design a resource allocation
algorithm which facilities power-efficient communication. Thus, we model the power dissipation in the system as the
sum of one static term and four dynamic terms as follows \cite{JR:Vijay_GBD}:\vspace*{-3mm}
\begin{eqnarray}\label{eqn:objective_function}
{\cal U}_{\mathrm{TP}}\Big(\mathbf{w}_k,s_l,P^{\mathrm{U}}_j\Big)=&&P_0+\underbrace{\sum_{l=1}^{N_{\mathrm{T}}L} s_l P^{\mathrm{Active}} +\sum_{l=1}^{N_{\mathrm{T}}L} (1-s_l) P^{\mathrm{Idle}}}_{\mbox{Antenna power consumption}}\notag\\
&&+  \underbrace{\eta\sum_{l=1}^{N_{\mathrm{T}}L} \sum_{k=1}^{K_{\mathrm{D}}} \varepsilon_{\mathrm{D}}\abs{{w}^l_k}^2 +\sum_{j=1}^{K_{\mathrm{U}}}\varepsilon_{\mathrm{U}}\zeta_j P^{\mathrm{U}}_j}_{\mbox{Amplifer power consumption}},
\end{eqnarray}
where $P_0$ is the aggregated static power consumption of the CP, all FD radio BSs, and all  backhaul links. $s_l\in\{0,1\}$ is a binary selection variable.  In particular, $s_l=1$ and $s_l=0$  indicate that the $l$-th antenna in  the FD communication system is in active mode and idle mode, respectively,  $s_l$ will be optimized to minimize the total network power consumption in the next section. $P^{\mathrm{Active}}>0$ is the signal processing power that is consumed if an antenna is active. $P^{\mathrm{Active}}$ includes   the power dissipations of the transmit filter, mixer, frequency synthesizer,  digital-to-analog converter, etc. In this paper, an FD radio antenna is considered active if it serves at least one user in the system. $P^{\mathrm{Idle}}>0$ is the required power consumption of an antenna in idle mode, i.e., if it is not serving any user, and $P^{\mathrm{Active}}> P^{\mathrm{Idle}}$ holds in general. $\sum_{k=1}^{K_{\mathrm{D}}}\sum_{l=1}^{N_{\mathrm{T}}L}\abs{{w}^l_k}^2 $ is the total power  radiated by the $L$ FD radio BSs for downlink transmission. $\varepsilon_{\mathrm{D}}\ge 1$ and
 $\varepsilon_{\mathrm{U}}\ge 1$ are constants which account for the inefficiency of the power amplifier \footnote{We assume  Class A power amplifiers with linear characteristic  are implemented in the transceivers. In practice, the maximum power efficiency of Class A amplifiers is $25\%$.} adopted for downlink and uplink transmission, respectively. In other words, $\varepsilon_{\mathrm{D}}\sum_{k=1}^{K_{\mathrm{D}}}\sum_{l=1}^{N_{\mathrm{T}}L}\abs{{w}^l_k}^2 $ and $\varepsilon_{\mathrm{U}}\sum_{j=1}^{K_{\mathrm{U}}} P^{\mathrm{U}}_j$ are the total power consumptions of the power amplifiers for downlink and uplink transmission, respectively. $\eta\ge 0$ and $\zeta_j\ge 0$ in the last two terms of (\ref{eqn:objective_function}) are constant weights which can be chosen by the system designer to prioritize the importance of the total downlink transmit power  and the transmit power of individual uplink users $j\in\{1,\ldots,K_{\mathrm{U}}\}$, respectively.
\vspace*{-3mm}
\section{Problem Formulation}
\label{sect:forumlation}
In this section, we first introduce the QoS
metrics for the considered FD radio communication network. Then, we formulate the resource allocation algorithm design as a non-convex optimization problem.
\subsection{Achievable Data Rate }
\label{subsect:Instaneous_Mutual_infxormation}
The  achievable data rate (bit/s/Hz) between the $L$ FD radio BSs and downlink user $k\in\{1,\ldots,K_{\mathrm{D}}\}$ is given by
\begin{eqnarray}
C_{k}=\log_2(1+\Gamma_{k}^{\mathrm{DL}}),\,\,
\mbox{where}\quad
\Gamma_{k}^{\mathrm{DL}}=\frac{\abs{\mathbf{h}_{\mathrm{D}_k}^H\mathbf{w}_k}^2}{\sum\limits_
{\substack{t\neq k}}^{K_{\mathrm{D}}}\abs{\mathbf{h}_{\mathrm{D}_k}^H\mathbf{w}_t}^2+\sum_{j=1}^{K_{\mathrm{U}}}P^{\mathrm{U}}_j\abs{g_{j,k}}^2+
\sigma_{\mathrm{n}_k}^2}
\end{eqnarray}
is the receive signal-to-interference-plus-noise ratio (SINR) at downlink user $k$.

On the other hand,  we assume that  the CP employs a  linear receiver for decoding of the received uplink information. Therefore,  the achievable data rate between the $L$ FD radio BSs and uplink user $j$  is given  by
\begin{eqnarray}\label{eqn:cap-eavesdropper}
C^{\mathrm{UL}}_j&=&\log_2\Big(1+\Gamma^{\mathrm{UL}}_j\Big),\quad \label{eqn:SINR_up_idle} \Gamma^{\mathrm{UL}}_j=\frac{P^{\mathrm{U}}_j\abs{\mathbf{v}_j^H\mathbf{h}_{\mathrm{U}_j}}^2 }{\sigma_\mathrm{z}^2 \norm{\mathbf{v}_j}^2+ I_j},\\
I_j&=&\mathbf{v}_j^H\Big(\sum_{k=1}^{K_{\mathrm{D}}} \mathbf{H}_{\mathrm{SI}} \mathbf{w}_k\mathbf{w}^H_k\mathbf{H}_{\mathrm{SI}}^H\Big) \mathbf{v}_j+\sum_{r\ne j}^{K_{\mathrm{U}}}P^{\mathrm{U}}_r\abs{\mathbf{v}_j^H\mathbf{h}_{\mathrm{U}_r}}^2 ,
   \end{eqnarray}
where $\mathbf{v}_j\in\mathbb{C}^{N_{\mathrm{T}}L\times 1}$ is the receive beamforming vector for decoding of the information for uplink user $j$. In this paper, maximum ratio combining (MRC) is adopted, i.e.,  the receive beamformer for uplink user $j$ is chosen as $\mathbf{v}_j=\sum_{l=1}^{N_{\mathrm{T}}L}s_l\mathbf{R}_l\mathbf{h}_{\mathrm{U}_j}$   to maximize the  signal strength of the received signal, where $\mathbf{R}_l\triangleq\diag\Big(\underbrace{0,\cdots,0}_{(l-1)},1,
\underbrace{0,\cdots,0}_{LN_\mathrm{T}-l}\Big),\forall l\in\{1,\ldots,L N_T\}$, is  a diagonal matrix.  It is known that   MRC  achieves a good system performance, especially if a large number of antennas is employed, and has been widely adopted in the literature \cite{JR:large_number_antennas,JR:TWC_large_antennas}.
 \begin{Remark}We note that zero-forcing beamforming (ZFBF) or minimum mean square error  beamforming (MMSE-BF) are not
considered for uplink signal detection since they do not facilitate an efficient  resource allocation algorithm design for the considered network.
 \end{Remark}
Using MRC, the uplink SINR of user $j$ is given by
\begin{eqnarray}\label{eqn:cap-eavesdropper-sm-sn1}
\Gamma^{\mathrm{UL}}_j=&&\frac{P^{\mathrm{U}}_j\Tr\Big(\mathbf{h}_{\mathrm{U}_j}\mathbf{h}_{\mathrm{U}_j}^H
\sum_{m=1}^{N_{\mathrm{T}}L}
\sum_{n=1}^{N_{\mathrm{T}}L}s_ms_n\mathbf{R}_m\mathbf{h}_{\mathrm{U}_j}
\mathbf{h}_{\mathrm{U}_j}^H\mathbf{R}_n^H\Big )}{ \sigma_{\mathrm{z}}^2\Tr\Big(\sum_{l=1}^{N_{\mathrm{T}}L}s_l\mathbf{h}_{\mathrm{U}_j}\mathbf{h}_{\mathrm{U}_j}^H\mathbf{R}_l\Big)+ I_j}, \quad\mbox{where}\label{eqn:cap-eavesdropper-sm-sn2}\\
I_j=&&\Tr\Big(\sum_{k=1}^{K_{\mathrm{D}}} \mathbf{H}_{\mathrm{SI}} \mathbf{w}_k\mathbf{w}^H_k\mathbf{H}_{\mathrm{SI}}^H\sum_{m=1}^{N_{\mathrm{T}}L}
\sum_{n=1}^{N_{\mathrm{T}}L}s_ms_n\mathbf{R}_m\mathbf{h}_{\mathrm{U}_j}
\mathbf{h}_{\mathrm{U}_j}^H\mathbf{R}_n^H\Big)\\
&&+ \sum_{r\ne j}^{K_{\mathrm{U}}}P^{\mathrm{U}}_r\Tr\Big(\mathbf{h}_{\mathrm{U}_r}\mathbf{h}_{\mathrm{U}_r}^H
\sum_{m=1}^{N_{\mathrm{T}}L}
\sum_{n=1}^{N_{\mathrm{T}}L}s_ms_n\mathbf{R}_m\mathbf{h}_{\mathrm{U}_j}
\mathbf{h}_{\mathrm{U}_j}^H\mathbf{R}_n^H\Big ).\label{eqn:cap-eavesdropper-sm-sn3}
   \end{eqnarray}

\subsection{Optimization Problem Formulation}
\label{sect:cross-Layer_formulation}
The system objective is to minimize  the total network power consumption while providing QoS for reliable communication to both uplink and downlink users simultaneously. We obtain the optimal resource allocation algorithm policy by  solving the following optimization problem:
\begin{eqnarray} \label{eqn:cross-layer}\notag
&&\hspace*{15mm} \underset{\mathbf{w}_k,s_l,P^{\mathrm{U}}_j}{\mino}\,\,\, {\cal U}_{\mathrm{TP}}\Big(\mathbf{w}_k,s_l,P^{\mathrm{U}}_j\Big)\\
\notag \mbox{s.t.}\hspace*{-6mm} &&\hspace*{2mm}\mbox{C1: }\frac{\abs{\mathbf{h}_{\mathrm{D}_k}^H\mathbf{w}_k}^2}{\sum\limits_
{\substack{t\neq k}}^{K_{\mathrm{D}}}\abs{\mathbf{h}_{\mathrm{D}_k}^H\mathbf{w}_t}^2+\sum_{j=1}^{K_{\mathrm{U}}}P^{\mathrm{U}}_j\abs{g_{j,k}}^2
+\sigma_{\mathrm{n}_k}^2}\ge\Gamma_{\mathrm{req}_k}^{\mathrm{DL}},\,\, \forall k\in\{1,\ldots,K_{\mathrm{D}}\}, \notag\\
\hspace*{-4mm}&&\hspace*{2mm}\mbox{C2: } \Gamma^{\mathrm{UL}}_j \ge\Gamma_{\mathrm{req}_j}^{\mathrm{UL}},\, \forall j\in\{1,\ldots,K_{\mathrm{U}}\},\notag\\
\hspace*{-4mm}&&\hspace*{2mm}\mbox{C3: }\sum_{k=1}^{K_{\mathrm{D}}}\abs{{w}^l_k}^2\le s_l P_{\max_l}^{\mathrm{DL}},\forall l\in\{1,\ldots,N_{\mathrm{T}}L\},\quad \mbox{C4: } 0\le P^{\mathrm{U}}_j\le P_{\max_j}^{\mathrm{U}},\forall j\in\{1,\ldots,K_{\mathrm{U}}\},\quad\notag\\
\hspace*{-4mm} &&\hspace*{2mm} \mbox{C5: }s_{l}\in\{0,1\},\forall l\in\{1,\ldots,N_{\mathrm{T}}L\}.
\end{eqnarray}
 $\Gamma_{\mathrm{req}_k}^{\mathrm{DL}}$  and $\Gamma_{\mathrm{req}_j}^{\mathrm{UL}}$ in constraints C1 and C2 denote the minimum  receive SINR required by downlink user $k$ and uplink user $j$ for successful information decoding, respectively.
 In C3, we constrain the maximum radiated power of the $l$-th antenna in the system to $P_{\max_l}^{\mathrm{DL}}$ to satisfy the maximum power spectral mask limit. C4 limits the maximum transmit power and ensures the non-negativity of the   transmit power of uplink user $j$.
 C5 constrains the  optimization variables which control the active and idle states of the antennas in the system to be binary.

\begin{Remark}
In this paper, energy/power saving is achieved by optimizing not only the uplink and downlink transmit powers, but also by optimizing the states of the antennas in the network. Thereby, it is expected that  switching the antennas on and off adaptively according to the channel conditions  is an effective strategy for reducing the network power consumption when the QoS requirements are not stringent or the number of users is low.
\end{Remark}

\section{Resource Allocation Algorithm Design} \label{sect:RA}
The optimization problem in (\ref{eqn:cross-layer})  is a mixed non-convex  and combinatorial optimization problem. The combinatorial nature is due to the binary  selection variables in C5. Also, variable $s_{l}$ is coupled with both downlink beamforming vector $\mathbf{w}_{k}$ and uplink power allocation variable $P^{\mathrm{U}}_j$ in constraint C2. Furthermore, constraint C1 is non-convex with respect to $\mathbf{w}_{k}$.   In the following, we first transform the optimization problem into an equivalent form and obtain the global optimal solution by using the generalized Bender's decomposition.  Then, we propose a  suboptimal  polynomial time algorithm which is inspired by the difference of convex functions program.
\vspace*{-3mm}
\subsection{Problem Reformulation}
In this section, we reformulate the considered optimization problem in (\ref{eqn:cross-layer}) using the definitions $\mathbf{W}_k=\mathbf{w}_k\mathbf{w}_k^H$, $\mathbf{H}_{\mathrm{D}_k}=\mathbf{h}_{\mathrm{D}_k}\mathbf{h}_{\mathrm{D}_k}^H$, and $\mathbf{H}_{\mathrm{
U}_j}=\mathbf{h}_{\mathrm{U}_j}\mathbf{h}_{\mathrm{U}_j}^H$. This leads to
\begin{eqnarray}\label{eqn:equivalent-binary}\notag
&&\hspace*{-10mm} \underset{\mathbf{W}_k\in\mathbb{H}^{N_{\mathrm{T}} L},s_l,P^{\mathrm{U}}_j,q_{m,n}}{\mino}\,\,\, P_0+\sum_{l=1}^{N_{\mathrm{T}}L} s_l P^{\mathrm{Active}} +\sum_{l=1}^{N_{\mathrm{T}}L} (1-s_l) P^{\mathrm{Idle}}+   \eta\sum_{k=1}^{K_{\mathrm{D}}}\varepsilon_{\mathrm{D}}\Tr(\mathbf{W}_k) +\varepsilon_{\mathrm{U}}\sum_{j=1}^{K_{\mathrm{U}}}\zeta_j P^{\mathrm{U}}_j\\
\notag \mbox{s.t.}\hspace*{-3mm} &&\hspace*{2mm}\mbox{C1: }\frac{\Tr(\mathbf{H}_{\mathrm{D}_k}\mathbf{W}_k)}{\Gamma_{\mathrm{req}_k}^{\mathrm{DL}}}\ge\sum\limits_
{\substack{t\neq k}}^{K_{\mathrm{D}}}\Tr(\mathbf{H}_{\mathrm{D}_k}\mathbf{W}_j)+\sum_{j=1}^{K_{\mathrm{U}}}P^{\mathrm{U}}_j\abs{g_{j,k}}^2+
\sigma_{\mathrm{n}_k}^2,\forall k, \\
\hspace*{-3mm}&&\hspace*{2mm}\notag\mbox{C2: } \frac{P^{\mathrm{U}}_j}{\Gamma_{\mathrm{req}_j}^{\mathrm{UL}}}\Tr\Big(\mathbf{H}_{\mathrm{U}_j}
\sum_{m=1}^{N_{\mathrm{T}}L}
\sum_{n=1}^{N_{\mathrm{T}}L}q_{m,n}\mathbf{R}_m\mathbf{H}_{\mathrm{U}_j}\mathbf{R}_n^H\Big ) \notag\\
\hspace*{-3mm}&&\hspace*{2mm}\ge\sigma_{\mathrm{z}}^2\Tr\Big(\sum_{l=1}^{N_{\mathrm{T}}L}q_{l,l}\mathbf{H}_{\mathrm{U}_j}\mathbf{R}_l\Big)+\Tr\Big(\sum_{k=1}^{K_{\mathrm{D}}} \mathbf{H}_{\mathrm{SI}} \mathbf{W}_k\mathbf{H}_{\mathrm{SI}}^H\sum_{m=1}^{N_{\mathrm{T}}L}
\sum_{n=1}^{N_{\mathrm{T}}L}q_{m,n}\mathbf{R}_m\mathbf{H}_{\mathrm{U}_j}
\mathbf{R}_n^H\Big)\notag\\
\hspace*{-3mm}&&\hspace*{2mm}+ \sum_{r\ne j}\Tr\Big(\mathbf{H}_{\mathrm{U}_r}
\sum_{m=1}^{N_{\mathrm{T}}L}
\sum_{n=1}^{N_{\mathrm{T}}L}P^{\mathrm{U}}_r q_{m,n}\mathbf{R}_m
\mathbf{H}_{\mathrm{U}_j}\mathbf{R}_n^H\Big )
,\, \forall j\in\{1,\ldots,K_{\mathrm{U}}\},\notag\\
\hspace*{-3mm}&&\hspace*{2mm}\notag\mbox{C3: }\sum_{k=1}^{K_{\mathrm{D}}} \Tr(\mathbf{W}_k\mathbf{R}_l)\le s_l P_{\max_l}^{\mathrm{DL}},\forall l\in\{1,\ldots,N_{\mathrm{T}}L\},\quad \mbox{C4,\,\, C5},\\
\hspace*{-3mm}&&\hspace*{2mm} \mbox{C6:}\,\, \mathbf{W}_k\succeq \mathbf{0},\,\, \forall k, \,
\mbox{C7:}\,\, \Rank(\mathbf{W}_k)\le 1,\,\, \forall k,  \mbox{C8:}\,\, 0\le q_{m,n}\le s_m,\,\, \forall m,n\in\{1,\ldots,N_{\mathrm{T}}L\},\notag\\
\hspace*{-3mm}&&\hspace*{2mm}\mbox{C9:}\,\, q_{m,n}\le s_n,\,\, \forall m,n,\quad
\mbox{C10:}\,\, q_{m,n}\ge s_n+s_m-1,\,\, \forall m,n.
\end{eqnarray}
 Constraints C6, C7, and $\mathbf{W}_k\in\mathbb{H}^{N_{\mathrm{T}} L},\forall k$, are imposed to guarantee that $\mathbf{W}_k=\mathbf{w}_k\mathbf{w}_k^H$ holds after optimization. $q_{m,n}$ is an auxiliary continuous optimization variable which is introduced to handle the product of two binary variables $s_ns_m$ in constraint C2, cf.  (\ref{eqn:cap-eavesdropper-sm-sn1})--(\ref{eqn:cap-eavesdropper-sm-sn3}).  In particular, because of  constraints C8 -- C10, $ q_{m,n}$ will have a binary value  if $s_{l}$ is binary.

We note that constraint C2 is still non-convex due to the product terms $q_{m,n}P_{t}^{\mathrm{U}}$ and $q_{m,n}\mathbf{W}_k$ which is an obstacle for the design of a computationally efficient resource allocation algorithm. In order to circumvent this difficulty, we adopt the big-M formulation  \cite{JR:big_M1,book:big_M} to decompose  the product terms.  First, we introduce auxiliary variables $\tilde{P}_{j,m,n}^{\mathrm{U}}={P}_{j}^{\mathrm{U}}q_{m,n}$ and $\widetilde{\mathbf{W}}_{k}^{m,n}={\mathbf{W}}_{k}q_{m,n}$. Then, we impose the following additional constraints:
\begin{subequations}
\begin{eqnarray}
\hspace*{-15mm}&&\hspace*{2mm}\mbox{C11: } \tilde{P}_{j,m,n}^{\mathrm{U}}\le  P_{\max_j}^{\mathrm{U}}q_{m,n}, \forall j,m,n, \hspace*{3cm} \mbox{C12: } \tilde{P}_{j,m,n}^{\mathrm{U}}\le {P}_{j}^{\mathrm{U}}, \forall j,m,n,\\
\hspace*{-15mm}&&\hspace*{2mm} \mbox{C13: } \tilde{P}_{j,m,n}^{\mathrm{U}}\ge  {P}_{j}^{\mathrm{U}}-(1-q_{m,n}) P_{\max_j}^{\mathrm{U}}, \forall j,m,n,\hspace*{0.90cm} \mbox{C14: } \tilde{P}_{j,m,n}^{\mathrm{U}}\ge  0,\\
\hspace*{-15mm}&&\hspace*{2mm} \mbox{C15: } \widetilde{\mathbf{W}}_{k}^{m,n}\preceq\mathbf{I}_{N_{\mathrm{T}}L}P_{\max_l}^{\mathrm{DL}}q_{m,n},  \forall k,m,n, \hspace*{2.2cm}\mbox{C16: } \widetilde{\mathbf{W}}_{k}^{m,n}\preceq{\mathbf{W}}_{k},  \forall k,m,n,\\
\hspace*{-15mm}&&\hspace*{2mm} \mbox{C17: } \widetilde{\mathbf{W}}_{k}^{m,n}\succeq{\mathbf{W}}_{k}-(1-q_{m,n})\mathbf{I}_{N_{\mathrm{T}}L}P_{\max_l}^{\mathrm{DL}},  \forall k,m,n, \hspace*{0cm}\mbox{C18: } \widetilde{\mathbf{W}}_{k}^{m,n}\succeq \zero,  \forall k,m,n.
\end{eqnarray}
\end{subequations}
In particular, constraints C11-C18 involve only continuous optimization variables, i.e., ${P}_{j}$, $\tilde{P}_{j,m,n}$, $q_{m,n}$, and $\mathbf{W}_{k}$,  which facilitates the design of an efficient resource allocation algorithm. Subsequently, we substitute $\tilde{P}_{j,m,n}^{\mathrm{U}}={P}_{j}^{\mathrm{U}}q_{m,n}$ and $\widetilde{\mathbf{W}}_{k}^{m,n}={\mathbf{W}}_{k}q_{m,n}$ into the coupled variables in C2 which yields\vspace*{-3mm}
 \begin{eqnarray}
\mbox{\textoverline{C2}: }
&& \frac{1}{\Gamma_{\mathrm{req}_j}^{\mathrm{UL}}}\Tr\Big(\mathbf{H}_{\mathrm{U}_j}
\sum_{m=1}^{N_{\mathrm{T}}L}
\sum_{n=1}^{N_{\mathrm{T}}L}\tilde{P}_{j,m,n}^{\mathrm{U}}\mathbf{R}_m\mathbf{H}_{\mathrm{U}_j}\mathbf{R}_n^H\Big ) \notag\\
\ge&&\sigma_{\mathrm{z}}^2\Tr\Big(\sum_{l=1}^{N_{\mathrm{T}}L}q_{l,l}\mathbf{H}_{\mathrm{U}_j}\mathbf{R}_l\Big)+
\Tr\Big(\sum_{k=1}^{K_{\mathrm{D}}}\sum_{m=1}^{N_{\mathrm{T}}L}
\sum_{n=1}^{N_{\mathrm{T}}L} \mathbf{H}_{\mathrm{SI}} \widetilde{\mathbf{W}}_{k}^{m,n}\mathbf{H}_{\mathrm{SI}}^H\mathbf{R}_m\mathbf{H}_{\mathrm{U}_j}
\mathbf{R}_n^H\Big)\notag\\
+ &&\sum_{r\ne j}\Tr\Big(\mathbf{H}_{\mathrm{U}_r}
\sum_{m=1}^{N_{\mathrm{T}}L}
\sum_{n=1}^{N_{\mathrm{T}}L}\tilde{P}_{r,m,n}^{\mathrm{U}}\mathbf{R}_m
\mathbf{H}_{\mathrm{U}_j}\mathbf{R}_n^H\Big )
,\, \forall j\in\{1,\ldots,K_{\mathrm{U}}\}.
\end{eqnarray}
The big-M formulation linearizes the terms ${q}_{m,n}P^{\mathrm{U}}_r$ and ${q}_{m,n}\mathbf{W}_k $ such that constraint $\mbox{\textoverline{C2} }$ is an affine function with respect to the new optimization variables $\tilde{P}^{\mathrm{U}}_{j,m,n}$ and $\widetilde{\mathbf{W}}_{k}^{m,n}$. We note that constraints C2 and $\mbox{\textoverline{C2} }$ are equivalent when constraints C5 and C11--C18 are satisfied.

As a result, the considered optimization problem (\ref{eqn:equivalent-binary}) can be transformed into the following equivalent problem:\vspace*{-2mm}
\begin{eqnarray}\label{eqn:equivalent-2}\notag
&&\hspace*{-10mm} \underset{\underset{P^{\mathrm{U}}_j,\tilde{P}^{\mathrm{U}}_{j,m,n}}{\mathbf{W}_k\in\mathbb{H}^{N_{\mathrm{T}} L},\widetilde{\mathbf{W}}_{k}^{m,n},s_l, q_{m,n},}}{\mino}\,\,\, P_0+\sum_{l=1}^{N_{\mathrm{T}}L} s_l P^{\mathrm{Active}} +\sum_{l=1}^{N_{\mathrm{T}}L} (1-s_l) P^{\mathrm{Idle}}+ \eta  \sum_{k=1}^{K_{\mathrm{D}}}\varepsilon_{\mathrm{D}}\Tr(\mathbf{W}_k) +\sum_{j=1}^{K_{\mathrm{U}}}\varepsilon_{\mathrm{U}}\zeta_j P^{\mathrm{U}}_j\\
\notag \mbox{s.t.}\hspace*{-2mm} &&\hspace*{12mm}\mbox{C1}, \mbox{\textoverline{C2}},  \mbox{C3, C4, C6}, \mbox{C8 -- C18},\\
\hspace*{-2mm} && \hspace*{12mm}\mbox{C5: }s_{l}\in\{0,1\},\forall l,\,\, \mbox{C7:}\,\, \Rank(\mathbf{W}_k)\le 1,\,\, \forall k, \end{eqnarray}
and we can focus on the design of an algorithm for solving the optimization problem  in (\ref{eqn:equivalent-2}). Now, the remaining non-convexity of optimization problem (\ref{eqn:equivalent-2}) is due to constraints C5 and C7.

\begin{Remark}
We note that the uplink-downlink duality approach in \cite{JR:Wei_Yu_duality,JR:UL_DL_dualtiy_CRN} cannot be applied to our problem for the following two reasons. First, the uplink and downlink transmit power variables  are coupled in constraints C1 and C2. Second, the  uplink and downlink transmit powers of each transceiver  are constrained.
\end{Remark}
\vspace*{-2mm}
\subsection{Optimal Iterative Resource Allocation Algorithm}
Now, we adopt the generalized Bender's decomposition (GBD) to handle the constraints involving binary optimization variables \cite{JR:Vijay_GBD}--\nocite{book:non_linear_and_mixed_integer,JR:generalized_Bender's}\cite{book:non_linear_integer}, i.e., C3, C8, C9, and C10. In particular, we decompose the problem in  (\ref{eqn:equivalent-2}) into two sub-problems: $(a)$ a \emph{primal problem} which is a non-convex optimization problem involving continuous optimization variables $\{\mathbf{W}_k,\widetilde{\mathbf{W}}_{k}^{m,n},P^{\mathrm{U}}_j,\tilde{P}^{\mathrm{U}}_{j,m,n},q_{m,n}\}$; $(b)$ a \emph{master problem} which is a mixed integer linear program (MILP). Specifically,  the primal problem
is solved for given $s_{l}$ which yields an upper bound for the optimal value of (\ref{eqn:equivalent-2}).  In contrast, the solution of the master problem  provides a lower bound for the optimal value of (\ref{eqn:equivalent-2}). Subsequently, we solve the primal and master problems iteratively until the solutions converge. In the following, we first propose algorithms for solving the primal and master problems in the $i$-th iteration, respectively. Then, we describe the iterative procedure between the master problem and the primal problem.

\subsubsection{Solution of the primal problem in the $i$-th iteration}
For given and fixed input parameters  $s_l=s_{l}(i)$ obtained from the master problem in the $i$-th iteration, we minimize the objective function with respect to variables $\{\mathbf{W}_k,\widetilde{\mathbf{W}}_{k}^{m,n},P^{\mathrm{U}}_j,\tilde{P}^{\mathrm{U}}_{j,m,n},q_{m,n}\}$ in the primal problem:
\begin{eqnarray}\label{eqn:primal_problem}
&&\hspace*{-15mm} \underset{\underset{P^{\mathrm{U}}_j,\tilde{P}^{\mathrm{U}}_{j,m,n}}{\mathbf{W}_k\in\mathbb{H}^{N_{\mathrm{T}} L},\widetilde{\mathbf{W}}_{k}^{m,n}, q_{m,n},}}{\mino}\,\, P_0+\sum_{l=1}^{N_{\mathrm{T}}L} s_l P^{\mathrm{Active}} +\sum_{l=1}^{N_{\mathrm{T}}L} (1-s_l) P^{\mathrm{Idle}}+ \eta  \sum_{k=1}^{K_{\mathrm{D}}}\varepsilon_{\mathrm{D}}\Tr(\mathbf{W}_k) +\varepsilon_{\mathrm{U}}\sum_{j=1}^{K_{\mathrm{U}}} \zeta_j P^{\mathrm{U}}_j\notag\\
\mbox{s.t.} &&\hspace*{15mm}\mbox{C1}, \mbox{\textoverline{C2}}, \mbox{C3, C4, C16 -- C18.}
\end{eqnarray}
We note that constraint C5  in (\ref{eqn:equivalent-2}) will be handled by the master problem since it involves only the binary optimization variable $s_{l}$. Now, the only obstacle in solving (\ref{eqn:primal_problem}) is the combinatorial rank constraint in C7 and we adopt the SDP relaxation approach to handle this non-convexity. In particular,  we relax constraint $\mbox{C7: }\Rank(\mathbf{W}_k)\le1$ by removing it from the problem formulation, such that the considered problem in (\ref{eqn:primal_problem}) becomes a convex SDP and can be solved efficiently by  numerical methods designed for convex programming such as interior point methods \cite{book:convex}. If the solution $\mathbf{W}_k$ of the relaxed version of (\ref{eqn:primal_problem}) is a rank-one matrix for all downlink users, then  the problem in (\ref{eqn:primal_problem}) and its relaxed version share the same optimal solution and the same optimal objective value.

Now, we study the tightness of the adopted SDP relaxation. The SDP relaxed  version of (\ref{eqn:primal_problem}) is jointly convex with respect to the optimization variables and satisfies Slater's constraint qualification. Thus, strong duality holds and  solving the dual problem is equivalent to solving (\ref{eqn:primal_problem}). To obtain the dual problem, we  define the Lagrangian
 of the relaxed version of (\ref{eqn:primal_problem}) as
\begin{eqnarray}
&&\hspace*{-7mm}{\cal L}\Big(\hspace*{-0.5mm}\mathbf{\Theta},\mathbf{\Phi}\hspace*{-0.5mm}\Big)={\cal U}_{\mathrm{TP}}\Big(\hspace*{-0.5mm}\mathbf{W}_k,s_l,
P^{\mathrm{U}}_j\hspace*{-0.5mm}\Big)\hspace*{-0.5mm}+\hspace*{-0.5mm}f_1(\mathbf{\Theta},\mathbf{\Phi}) \hspace*{-0.5mm}+\hspace*{-0.5mm} f_2(\mathbf{\Theta},\mathbf{\Phi}),\, \quad \mbox{where}\\
&&\hspace*{-7mm}\label{eqn:f0}{\cal U}_{\mathrm{TP}}\Big(\hspace*{-0.5mm}\mathbf{W}_k,s_l,P^{\mathrm{U}}_j\hspace*{-0.5mm}\Big)=P_0\hspace*{-0.5mm}+\hspace*{-0.5mm}\sum_{l=1}^{N_{\mathrm{T}}L} s_l P^{\mathrm{Active}} \hspace*{-0.5mm}+\hspace*{-0.5mm}\sum_{l=1}^{N_{\mathrm{T}}L} (1-s_l) P^{\mathrm{Idle}}+ \eta \sum_{k=1}^{K_{\mathrm{D}}}\varepsilon_{\mathrm{D}}\Tr(\mathbf{W}_k)\hspace*{-0.5mm} +\hspace*{-0.5mm}\varepsilon_{\mathrm{U}}\sum_{j=1}^{K_{\mathrm{U}}}\zeta_jP^{\mathrm{U}}_j\\
&&\hspace*{-7mm}\label{eqn:f1}f_1(\mathbf{\Theta},\mathbf{\Phi}) =-\sum_{k=1}^{K_{\mathrm{D}}}\Tr(\mathbf{Z}_k\mathbf{W}_k)
-
\sum_{j=1}^{K_{\mathrm{U}}}\sum_{m=1}^{N_{\mathrm{T}}L}\sum_{n=1}^{N_{\mathrm{T}}L}\beta_{j,m,n} \tilde{P}_{j,m,n}^{\mathrm{U}}+\sum_{j=1}^{K_{\mathrm{U}}}\lambda_j(P^{\mathrm{U}}_{j}- P_{\max_j}^{\mathrm{U}})\notag\\
&&\hspace*{-7mm}+
\sum_{k=1}^{K_{\mathrm{D}}}\alpha_k\Big[-\frac{\Tr(\mathbf{H}_{\mathrm{D}_k}\mathbf{W}_k)}{\Gamma_{\mathrm{req}_k}^{\mathrm{DL}}}+\sum\limits_
{\substack{t\neq k}}^{K_{\mathrm{D}}}\hspace*{-0.5mm}\Tr(\mathbf{H}_{\mathrm{D}_k}\mathbf{W}_j)+\sum_{j=1}^{K_{\mathrm{U}}}P^{\mathrm{U}}_j\abs{g_{j,k}}^2+
\sigma_{\mathrm{z}}^2\Big]-\sum_{j=1}^{K_{\mathrm{U}}}\chi_jP^{\mathrm{U}}_{j}\notag\\
&&\hspace*{-7mm}
+\sum_{j=1}^{K_{\mathrm{U}}}\psi_j\Bigg(\frac{-\Tr\Big(\mathbf{H}_{\mathrm{U}_j}
\sum_{m=1}^{N_{\mathrm{T}}L}
\sum_{n=1}^{N_{\mathrm{T}}L}\tilde{P}_{j,m,n}^{\mathrm{U}}\mathbf{R}_m\mathbf{H}_{\mathrm{U}_j}\mathbf{R}_n^H\Big ) }{\Gamma_{\mathrm{req}_j}^{\mathrm{UL}}} +\sigma_{\mathrm{z}}^2\Tr\Big(\sum_{l=1}^{N_{\mathrm{T}}L}q_{l,l}\mathbf{H}_{\mathrm{U}_j}\mathbf{R}_l\Big)\notag\\
&&\hspace*{-7mm}+\Tr\Big(\sum_{k=1}^{K_{\mathrm{D}}}\sum_{m=1}^{N_{\mathrm{T}}L}
\sum_{n=1}^{N_{\mathrm{T}}L} \mathbf{H}_{\mathrm{SI}} \widetilde{\mathbf{W}}_{k}^{m,n}\mathbf{H}_{\mathrm{SI}}^H\mathbf{R}_m\mathbf{H}_{\mathrm{U}_j}
\mathbf{R}_n^H\Big)+ \sum_{r\ne j}\Tr\Big(\mathbf{H}_{\mathrm{U}_r}
\sum_{m=1}^{N_{\mathrm{T}}L}
\sum_{n=1}^{N_{\mathrm{T}}L}\tilde{P}_{r,m,n}^{\mathrm{U}}\mathbf{R}_m
\mathbf{H}_{\mathrm{U}_j}\mathbf{R}_n^H\Big )\Bigg)\notag\\
&&\hspace*{-7mm} + \sum_{j=1}^{K_{\mathrm{U}}}\sum_{m=1}^{N_{\mathrm{T}}L} \sum_{n=1}^{N_{\mathrm{T}}L} \mu_{j,m,n} (\tilde{P}^{\mathrm{U}}_{j,m,n}-P_{\max_j}^{\mathrm{U}}q_{m,n})+\sum_{j=1}^{K_{\mathrm{U}}}\sum_{m=1}^{N_{\mathrm{T}} L}\sum_{n=1}^{N_{\mathrm{T}} L}\xi_{j,m,n}\Big(
{P}_{j}^{\mathrm{U}}-(1-q_{m,n}) P_{\max_j}^{\mathrm{U}}-\tilde{P}_{j,m,n}^{\mathrm{U}}\Big)\notag
\\
&&\hspace*{-7mm} + \sum_{j=1}^{K_{\mathrm{U}}}\sum_{m=1}^{N_{\mathrm{T}}L} \sum_{n=1}^{N_{\mathrm{T}}L} \tau_{j,m,n} (\tilde{P}_{j,m,n}^{\mathrm{U}}- {P}_{j}^{\mathrm{U}})-\sum_{m=1}^{N_{\mathrm{T}}L}\sum_{n=1}^{N_{\mathrm{T}}L}\varsigma_{m,n} q_{m,n}\notag\\
&&\hspace*{-7mm}+\sum_{k=1}^{K_{\mathrm{D}}}\sum_{m=1}^{N_{\mathrm{T}} L}\sum_{n=1}^{N_{\mathrm{T}} L}\Tr\Bigg\{\mathbf{D}_{\mathrm{C}_{15_{k,m,n}}}\Big(\widetilde{\mathbf{W}}_{k}^{m,n}-
\mathbf{I}_{N_{\mathrm{T}}L}P_{\max_l}^{\mathrm{DL}}q_{m,n}\Big)+\mathbf{D}_{\mathrm{C}_{16_{k,m,n}}}\Big(\widetilde{\mathbf{W}}_{k}^{m,n}
-{\mathbf{W}}_{k}\Big)\Bigg\}\notag\\
&&\hspace*{-7mm}+\sum_{k=1}^{K_{\mathrm{D}}}\sum_{m=1}^{N_{\mathrm{T}} L}\sum_{n=1}^{N_{\mathrm{T}} L}\Tr\Bigg\{\mathbf{D}_{\mathrm{C}_{17_{k,m,n}}}\Big({\mathbf{W}}_{k}\hspace*{-0.5mm}-\hspace*{-0.5mm}(1-q_{m,n})\mathbf{I}_{N_{\mathrm{T}}L}
P_{\max_l}^{\mathrm{DL}}\hspace*{-0.5mm}-\hspace*{-0.5mm}
\widetilde{\mathbf{W}}_{k}^{m,n}
\Big)\hspace*{-0.5mm}-\hspace*{-0.5mm}\mathbf{D}_{\mathrm{C}_{18_{k,m,n}}}\widetilde{\mathbf{W}}_{k}^{m,n}
\Bigg\}, \mbox{and}\\
&&\hspace*{-7mm}\label{eqn:f2}\notag
f_2(\mathbf{\Theta},\mathbf{\Phi})=\sum_{k=1}^{K_{\mathrm{D}}}\sum_{l=1}^{N_{\mathrm{T}}L} \rho_{l}\Big(\hspace*{-0.5mm}\Tr(\mathbf{W}_k\mathbf{R}_l)- s_l P_{\max_l}^{\mathrm{DL}}\hspace*{-0.5mm}\Big) + \sum_{m=1}^{N_{\mathrm{T}}L} \sum_{n=1}^{N_{\mathrm{T}}L} \kappa_{m,n} (q_{m,n}- s_m)\\
&&\hspace*{-7mm}+\sum_{m=1}^{N_{\mathrm{T}}L} \sum_{n=1}^{N_{\mathrm{T}}L} \varphi_{m,n} (q_{m,n}- s_n)+\sum_{m=1}^{N_{\mathrm{T}}L} \sum_{n=1}^{N_{\mathrm{T}}L} \omega_{m,n} ( s_n+s_m-1-q_{m,n}).
\end{eqnarray}
Here, $\mathbf{\Theta}=\{\mathbf{W}_k,s_l,P^{\mathrm{U}}_j,\widetilde{\mathbf{W}}_{k}^{m,n},\tilde{P}^{\mathrm{U}}_{j,m,n},q_{m,n}\}$ and
$\boldsymbol \Phi=\{\alpha_k,\psi_j,\rho_l,\{\lambda_j,\chi_j\},\mathbf{Z}_k,\{\varsigma_{m,n} ,\kappa_{m,n}\},\varphi_{m,n},$ $\omega_{m,n},$ $\mu_{j,m,n}
\tau_{j,m,n},
\xi_{j,m,n},\beta_{j,m,n}, \mathbf{D}_{\mathrm{C}_{15_{k,m,n}}},\mathbf{D}_{\mathrm{C}_{16_{k,m,n}}},\mathbf{D}_{\mathrm{C}_{17_{k,m,n}}},
\mathbf{D}_{\mathrm{C}_{18_{k,m,n}}}\}$ are the collections of primal and dual variables, respectively; $ \alpha_k\ge0,\psi_j\ge0,\rho_l\ge0,\{\lambda_j,\chi_j\}\ge0,\mathbf{Z}_k\succeq \zero,\varsigma_{m,n} \ge 0,\kappa_{m,n}\ge0,\varphi_{m,n}\ge0,\omega_{m,n}\ge0,\mu_{j,m,n}\ge0$,
$\tau_{j,m,n}\ge0,
\xi_{j,m,n}\ge0,\beta_{j,m,n}\ge0, \mathbf{D}_{\mathrm{C}_{15_{k,m,n}}}\succeq \zero,\mathbf{D}_{\mathrm{C}_{16_{k,m,n}}}\succeq \zero,\mathbf{D}_{\mathrm{C}_{17_{k,m,n}}}\succeq \zero,$ and $
\mathbf{D}_{\mathrm{C}_{18_{k,m,n}}}\succeq \zero$, are the scalar/matrix dual variables for constraints  C1 -- C4, C8 -- C18, respectively. Function ${\cal U}_{\mathrm{TP}}\Big(\mathbf{W}_k,s_l,P^{\mathrm{U}}_j\Big)$  in (\ref{eqn:f0}) is the objective function of the SDP relaxed version of problem (\ref{eqn:primal_problem}); $f_1(\mathbf{\Theta},\mathbf{\Phi}) $ in (\ref{eqn:f1}) is a function involving the constraints that do not depend on the binary optimization variables;
$f_2(\mathbf{\Theta},\mathbf{\Phi})$ in (\ref{eqn:f2}) is a function involving the constraints including $s_{l}(i)$. These functions are introduced here for notational simplicity and will be exploited for facilitating the presentation of the solutions for both the primal problem and the master problem.

For a given $s_l$, the dual problem of the SDP relaxed optimization problem in (\ref{eqn:primal_problem}) is given by
\begin{equation}\hspace*{-10mm}\label{eqn:dual}
\underset{\boldsymbol \Phi}{\maxo} \,\,\underset{\mathbf{\Theta}}{\mino} \,{\cal L}\Big(\mathbf{\Theta},\mathbf{\Phi}\Big).
\end{equation}
 We define $\boldsymbol \Theta^*(i)=\{\mathbf{W}_k^*,s_l,P^{\mathrm{U}*}_j,\widetilde{\mathbf{W}}_{k}^{m,n*},\tilde{P}^{\mathrm{U}*}_{j,m,n},q_{m,n}^*\}$ and $\boldsymbol \Phi(i)=\{\boldsymbol \Phi^* \}$ as the optimal primal solution and the optimal dual solution of the SDP relaxed problem in (\ref{eqn:primal_problem}) in the $i$-th iteration.

Now, we introduce the following theorem regarding the tightness of the adopted SDP relaxation.
\begin{Thm}\label{thm:rankone_condition} Assuming the channel vectors of the downlink users, $\mathbf{h}_{\mathrm{D}_k},k\in\{1,\ldots,K_{\mathrm{D}}\},$ can be modeled as statistically independent  random variables, then the solution of the SDP relaxed version of (\ref{eqn:primal_problem}) is rank-one, i.e.,  $\Rank(\mathbf{W}_k)=1,\forall k$, with probability one.  Thus, the optimal downlink beamformer for user $k$, i.e., $\mathbf{w}_k$, is  the principal eigenvector of $\mathbf{W}_k$.
\end{Thm}
\emph{\quad Proof: } Please refer to Appendix A for a proof of Theorem $1$.

On the other hand,  we formulate an $l_1$-minimization problem for the case when (\ref{eqn:primal_problem}) is infeasible for given binary variables $s_{l}(i)$. The $l_1$-minimization problem is given as:
 \begin{eqnarray}\label{eqn:FP} \label{eqn:l-1_problem}
\hspace*{-5mm}&& \hspace*{-5mm}\underset{\underset{P^{\mathrm{U}}_j,\tilde{P}^{\mathrm{U}}_{j,m,n},\nu_{l}^{\mathrm{C3}},
\nu_{m,n}^{\mathrm{C8}},\nu_{m,n}^{\mathrm{C9}},\nu_{m,n}^{\mathrm{C10}}}{\mathbf{W}_k\in\mathbb{H}^{N_{\mathrm{T}} L},\widetilde{\mathbf{W}}_{k}^{m,n}, q_{m,n},}}{\mino}\,\, \sum_{l=1}^{N_{\mathrm{T}}L} \nu_{l}^{\mathrm{C3}}+ \sum_{m=1}^{N_{\mathrm{T}}L}\sum_{n=1}^{N_{\mathrm{T}}L}\nu_{m,n}^{\mathrm{C8}}+\
\nu_{m,n}^{\mathrm{C9}}+\nu_{m,n}^{\mathrm{C10}}\\
\hspace*{-8mm} &&\mbox{s.t.}\hspace*{20mm}\mbox{C1}, \mbox{\textoverline{C2}}, \mbox{C4, C6, C11 -- C18},\notag \\
\hspace*{-8mm}&&\hspace*{5mm}\notag\mbox{C3: } \Tr(\mathbf{W}_k\mathbf{R}_l)\le s_{l}(i) P_{\max_l}^{\mathrm{DL}} + \nu_{l}^{\mathrm{C3}},\forall l\in\{1,\ldots,N_{\mathrm{T}}L\},\\
\hspace*{-8mm}&&\hspace*{5mm} \mbox{C8:}\,\, 0\le q_{m,n}\le s_m(i)+\nu_{m,n}^{\mathrm{C8}},\,\, \forall m,n\in\{1,\ldots,N_{\mathrm{T}}L\},\notag\,\mbox{C9:}\,\, q_{m,n}\le s_n(i)+\nu_{m,n}^{\mathrm{C9}},\,\, \forall m,n,\\
\hspace*{-8mm}&&\hspace*{5mm}\mbox{C10:}\,\, \nu_{m,n}^{\mathrm{C10}}+q_{m,n}\ge s_n(i)+s_m(i)-1,\,\, \forall m,n,\,\mbox{C19:}\,\, \nu_{l}^{\mathrm{C3}},\nu_{m,n}^{\mathrm{C8}},\nu_{m,n}^{\mathrm{C9}},\nu_{m,n}^{\mathrm{C10}}\ge 0, \forall l,m,n,k.\notag
\end{eqnarray}
Equation (\ref{eqn:l-1_problem}) is an SDP problem and can be solved by  interior point methods with polynomial time computational complexity. We note that the objective function in (\ref{eqn:l-1_problem}) is the sum of the  constraint violations with respect to the problem in (\ref{eqn:primal_problem}).  Besides, the corresponding dual variables and the optimal primal variables will be used as the input to the master problem for the next iteration \cite{book:non_linear_and_mixed_integer}. We adopt a similar notation as in (\ref{eqn:equivalent-binary})  to denote the primal and dual variables in (\ref{eqn:FP}). In particular,  the
primal and dual solutions for the $l_1$-minimization problem in (\ref{eqn:FP}) are denoted as
$\mathbf{\overline \Theta}=\{\mathbf{\overline W}_k,\overline s_l, \overline P^{\mathrm{U}}_j,{\widetilde{\mathbf{W}}_{k}^{m,n}},{\tilde{P}^{\mathrm{U}}_{j,m,n}},{q_{m,n}}\}$ and
$\boldsymbol {\overline\Phi}=\{\overline\alpha_k,\overline\psi_j,\overline\rho_l,\{\overline\lambda_j,\overline\chi_j\},\mathbf{\overline Z}_k,\overline\varsigma_{m,n},\overline\kappa_{m,n},\overline\varphi_{m,n},\overline\omega_{m,n},$ $\overline\mu_{j,m,n},$ $
\overline\tau_{j,m,n},\overline\xi_{j,m,n},\overline\beta_{j,m,n}, \mathbf{\overline D}_{\mathrm{C}_{15_{k,m,n}}},\mathbf{\overline D}_{\mathrm{C}_{16_{k,m,n}}},$
$\mathbf{\overline D}_{\mathrm{C}_{17_{k,m,n}}},
\mathbf{\overline D}_{\mathrm{C}_{18_{k,m,n}}}\}$, respectively. The primal and dual variables will be exploited as inputs for the constraints of the master problem.

\subsubsection{Solution of the master problem in the $i$-th iteration}
For notational simplicity, we define $\cal F$ and $\cal I$ as the sets of all iteration indices at which the primal problem is feasible and infeasible, respectively. Then, we formulate the master problem which utilizes the solutions of (\ref{eqn:equivalent-binary}) and (\ref{eqn:FP}). The master problem in the $i$-th iteration is given as follows:
\begin{subequations}\label{eqn:master_problem}\notag
\begin{eqnarray}\label{eqn:master_problem_objective}
&&\hspace*{10mm}\underset{\varrho,\, s_{l}}{\mino}\,\, \varrho\\
\hspace*{15mm}\mbox{s.t.} &&\hspace*{0mm}\mbox{C5},\\
&&\varrho \ge  \xi(\boldsymbol \Phi(t), s_{l}),   t\in\{1,\ldots,i\} \cap   \cal F,\label{eqn:supporting_plane_constraint1}\\
&&0\ge  \overline{ \xi}(\boldsymbol {\overline\Phi}(t), s_{l}),  t\in\{1,\ldots,i\} \cap  \cal I,\label{eqn:supporting_plane_constraint2}
\end{eqnarray}
\end{subequations}
where $s_{l}$ and $\varrho$ are optimization variables for the master problem and
\begin{eqnarray}\label{eqn:opt_master_1}\hspace*{-2.5mm}
\xi( \boldsymbol{\Phi}(t),s_l)\hspace*{-1.5mm}&=&\hspace*{-1.5mm} \underset{\underset{P^{\mathrm{U}}_j,\tilde{P}^{\mathrm{U}}_{j,m,n}}{\mathbf{W}_k\in\mathbb{H}^{N_{\mathrm{T}} L},\widetilde{\mathbf{W}}_{k}^{m,n}, q_{m,n},}}{\mino}\,\, {\cal U}_{\mathrm{TP}}\Big(\hspace*{-0.5mm}\mathbf{W}_k,s_l,P^{\mathrm{U}}_j\hspace*{-0.5mm}\Big) +f_1(\mathbf{\Theta},\mathbf{\Phi}(t))+ f_2(\mathbf{\Theta},\mathbf{\Phi}(t)),\\
\hspace*{-2.5mm}\overline\xi(\boldsymbol {\overline\Phi}(t),s_l)\hspace*{-1.5mm}&=& \hspace*{-1.5mm}\underset{\underset{P^{\mathrm{U}}_j,\tilde{P}^{\mathrm{U}}_{j,m,n}}{\mathbf{W}_k\in\mathbb{H}^{N_{\mathrm{T}} L},\widetilde{\mathbf{W}}_{k}^{m,n}, q_{m,n},}}{\mino}\,\,f_1(\boldsymbol{\overline \Theta},\boldsymbol {\overline\Phi}(t))+ f_2(\boldsymbol{\overline \Theta},\boldsymbol {\overline\Phi}(t))  .\label{eqn:opt_master_2}
\end{eqnarray}
Equations (\ref{eqn:opt_master_1}) and (\ref{eqn:opt_master_2}) are two different minimization problems defining the constraint set of  the master problem in (\ref{eqn:master_problem}).  In particular, $\varrho\ge \xi( \boldsymbol{\Phi}(t),s_l), t\in\{1,\ldots,i\} \cap  \cal F$ in (\ref{eqn:supporting_plane_constraint1})  and $0\ge\overline\xi(\boldsymbol {\overline\Phi}(t),s_l), t\in\{1,\ldots,i\} \cap  \cal I$  in (\ref{eqn:supporting_plane_constraint2}) denote the sets of hyperplanes spanned by the \emph{optimality cut} and the \emph{feasibility cut} from the first to the $i$-th iteration, respectively. The two different types of hyperplanes reduce the search region for the global optimal solution. Moreover, both $\xi( \boldsymbol{\Phi}(t),s_l)$ and $\overline\xi(\boldsymbol {\overline\Phi}(t),s_l)$ are also functions  of $s_{l}$ which is the optimization variable of the outer minimization in (\ref{eqn:master_problem}).

Now, we introduce the following proposition for the solution of the two minimization problems in (\ref{eqn:opt_master_1}) and (\ref{eqn:opt_master_2}).

\begin{Prop}\label{thm:supporting_plane}
The solutions of (\ref{eqn:opt_master_1}) and (\ref{eqn:opt_master_2}) for index $t\in\{1,\ldots,i\}$  are  the solutions of (\ref{eqn:primal_problem}) and (\ref{eqn:FP}) in the $t$-th iteration, respectively.
\end{Prop}

\emph{\quad Proof: } Please refer to Appendix B for a proof of Proposition \ref{thm:supporting_plane}.

The master problem in (\ref{eqn:master_problem}) is transformed to an MILP by applying Proposition \ref{thm:supporting_plane} to solve (\ref{eqn:opt_master_1}) and (\ref{eqn:opt_master_2}). Hence, the master problem can be solved by using standard numerical solvers for MILPs such as Mosek \cite{JR:Mosek} and Gurobi \cite{JR:Gurobi}. We note that an additional constraint is imposed to the master problem in each additional
 iteration, thus the objective value of (\ref{eqn:master_problem}), i.e., the lower bound of (\ref{eqn:equivalent-2}), is a monotonically non-decreasing function with respect to the number of iterations.
\subsubsection{Overall algorithm}
\begin{table}[t]\vspace*{-1.5cm}\caption{Optimal Iterative Resource Allocation Algorithm based on GBD\vspace*{-0.8cm}}\label{table:algorithm}

\renewcommand\thealgorithm{}
\begin{algorithm} [H]                
\caption{Generalized Bender's Decomposition}          
\label{alg1}                           
\begin{algorithmic} [1]
\STATE Initialize the maximum number of iterations $L_{\max}$, UB$(0)=\infty$, LB$(0)=-\infty$, and a small constant $\vartheta\rightarrow 0$
\STATE Set iteration index $i=1$ and start with   $s_{l}(i)=1,\forall k,l$

\REPEAT [Loop]


\STATE Solve  (\ref{eqn:primal_problem}) by SDP relaxation for a given set of $s_{l}(i)$

\IF{(\ref{eqn:primal_problem}) is feasible}

\STATE Obtain an intermediate resource allocation policy $\boldsymbol \Theta'(i)=\{\mathbf{W}_k',s_l,P^{\mathrm{U}'}_j,\widetilde{\mathbf{W}}_{k}^{m,n'},\tilde{P}^{\mathrm{U}'}_{j,m,n},{q}_{m,n}'\}$, the corresponding Lagrange multiplier set $\boldsymbol\Phi(i)$, and an intermediate objective value $f_0'$

\STATE The upper bound value is updated with $\mathrm{UB}(i)={\min} \{\mathrm{UB}(i-1), f_0'\} $. If $\mathrm{UB}(i)=f_0'$, we set the current optimal policy $\boldsymbol \Theta_{\mathrm{current}}=\boldsymbol\Theta(i)$

\ELSE
\STATE  Solve the $l_1$-minimization problem in (\ref{eqn:FP}) and obtain an intermediate resource allocation policy $\mathbf{\widetilde \Theta}(i)=\{\mathbf{W}_k',s_l,P^{\mathrm{U}'}_j,\widetilde{\mathbf{W}}_{k}^{m,n'},\tilde{P}^{\mathrm{U}'}_{j,m,n},{q}_{m,n}'\}$ and the corresponding Lagrange multiplier set  $\boldsymbol{\widetilde\Phi}(i)$
 \ENDIF

\STATE Solve the master problem in (\ref{eqn:master_problem}) for $s_{l}$, save $s_{l}(i+1)=s_{l}$, and obtain the $i$-th lower bound, i.e., $\mathrm{LB}(i)$

\IF{$\abs{\mathrm{LB}(i)-\mathrm{UB}(i)}\le \vartheta$}
\STATE
Global optimal = \TRUE, \textbf{return} $\{\mathbf{W}_k^*,s_l^*,P^{\mathrm{U}*}_j,\widetilde{\mathbf{W}}_{k}^{m,n*},\tilde{P}^{\mathrm{U}*}_{j,m,n},q_{m,n}^*\}=\{\boldsymbol \Theta_{\mathrm{current}}\}$
\ELSE
\STATE $i=i+1$
 \ENDIF
\UNTIL{ $i=L_{\max}$}

\end{algorithmic}
\end{algorithm}
\vspace*{-1.7cm}
\end{table}
The proposed iterative resource allocation algorithm is summarized in Table \ref{table:algorithm} and is implemented by a repeated loop.  For the initiation, we first set the iteration index $i$ to one and  the binary variables $s_{l}(i)$ to one, e.g. $s_{l}(1)=1, \forall l$. In the $i$-th iteration, we solve the problem in (\ref{eqn:primal_problem}) via SDP relaxation.  Two different types of Lagrange multipliers  are
defined depending on the feasibility of the primal problem. If the problem is feasible  for a given $s_{l}(i)$ (lines 6, 7),  then we obtain an intermediate resource allocation policy $\mathbf{\Theta}(i)$, an intermediate objective value $f_0'$, and the corresponding Lagrange multiplier set $\boldsymbol\Phi(i)$. In particular, $\boldsymbol\Phi(i)$ is used to generate an \emph{optimality cut} in the master problem. Also,  the optimal resource allocation policy and the performance upper bound $\mathrm{UB}(i)$  are updated if the computed objective value is the lowest across all the iterations. On the contrary,  if the primal problem is infeasible for a given $s_{l}(i)$ (line 9), then we solve the $l_1$-minimization problem in (\ref{eqn:FP}) and obtain an intermediate resource allocation policy $\mathbf{\widetilde \Theta}(i)$ and the corresponding Lagrange multiplier set $\boldsymbol {\overline\Phi}(i)$. This information will be used to generate an \emph{infeasibility  cut} in the master problem. We note that the
upper bound is obtained only from the feasible primal problem.  Subsequently, we solve the master problem  based on $\mathbf{\widetilde \Theta}(t)$ and $\mathbf{\Theta}(i)$, $t\in\{1,\ldots,i\}$,   via a standard MILP numerical solver. Due to weak duality \cite{book:non_linear_integer}, the optimal value of the original optimization problem in (\ref{eqn:primal_problem}) is bounded below by the objective value of the master problem in each iteration. The algorithm stops when the difference between the $i$-th lower bound and the $i$-th upper bound is smaller than a predefined threshold $\vartheta\ge 0$ (lines 12 -- 14).  We note that when the master and the primal problems can be solved in each iteration, the proposed algorithm is guaranteed to converge to the optimal solution  \cite[Theorem 6.3.4]{book:non_linear_and_mixed_integer}.
%
\vspace*{-2mm}
\subsection{Suboptimal Resource Allocation Algorithm Design}
The optimal iterative resource allocation algorithm proposed  in the last section  has a non-polynomial time computational complexity due to the MILP master problem \footnote{The optimal algorithm serves mainly as a performance benchmark for the proposed suboptimal algorithm.}. In this section, we propose a suboptimal resource allocation algorithm which has a polynomial time computational complexity. The starting point for the design of the proposed suboptimal resource allocation algorithm is  the reformulated optimization problem in (\ref{eqn:equivalent-binary}).
\subsubsection{Problem reformulation via difference of convex functions programming}
The major obstacle in solving (\ref{eqn:equivalent-binary}) are the binary constraints. Hence, we rewrite constraint C5 in its equivalent form:
\begin{eqnarray}
\hspace*{-8mm}\mbox{C5a: }&& \hspace*{-4mm}0 \le s_{l}\le 1,\forall l\in\{1\ldots,L\} \quad \mbox{and} \hspace*{25mm} \mbox{C5b: } \sum_{l=1}^{N_{\mathrm{T}}L}   s_{l} -\sum_{l=1}^{N_{\mathrm{T}}L} s_{l}^2 \le 0.
\end{eqnarray}
Now, optimization variables $ s_{l}$  in C5a are continuous values between zero and one while constraint  C5b is the difference of two convex functions. By using the SDP relaxation approach as in the optimal resource allocation algorithm, we can reformulate the optimization problem as
 \begin{eqnarray}\label{eqn:equivalent-dc-constraint}
&&\hspace*{-15mm} \underset{\mathbf{W}_k\in \mathbb{H}^{N_{\mathrm{T}}L},\widetilde{\mathbf{W}}_{k,b}^l,P^{\mathrm{U}}_j,\tilde{P}^{\mathrm{U}}_{j,m,n},q_{m,n}}{\mino}\,\, {\cal U}_{\mathrm{TP}}\Big(\hspace*{-0.5mm}\mathbf{W}_k,s_l,P^{\mathrm{U}}_j\hspace*{-0.5mm}\Big)\\
\mbox{s.t.} &&\hspace*{15mm}\mathbf{\Theta}\in{\cal D},\,\, \mbox{C5b},\notag
\end{eqnarray}
where ${\cal D}$ denotes the  convex feasible  solution set spanned by constraints $\mbox{C1}, \mbox{\textoverline{C2}}, \mbox{C3, C4, C5a, C6},$  and $\mbox{ C8 -- C18.}$
The only non-convexity in (\ref{eqn:equivalent-dc-constraint}) is due to constraint C5b which is a reverse convex function \cite{JR:DC_programming}. Now, we introduce the following Theorem for handling the constraint.
 \begin{Thm}\label{Thm:penalty_method}
 For a large constant value $\phi\gg1$,   (\ref{eqn:equivalent-dc-constraint}) is equivalent\footnote{Here, equivalence means that both problems share the same optimal objective value and the same optimal resource allocation policy.} to  the following problem:
 \begin{eqnarray}\label{eqn:equivalent-approx2}
&& \hspace*{-15mm} \underset{\mathbf{W}_k\in \mathbb{H}^{N_{\mathrm{T}}L},\widetilde{\mathbf{W}}_{k,b}^l,P^{\mathrm{U}}_j,\tilde{P}^{\mathrm{U}}_{j,m,n},q_{m,n}}{\mino}\,\, {\cal U}_{\mathrm{TP}}\Big(\hspace*{-0.5mm}\mathbf{W}_k,s_l,P^{\mathrm{U}}_j\hspace*{-0.5mm}\Big)+\phi\Big(\sum_{l=1}^{N_{\mathrm{T}}L}   s_{l} -\sum_{l=1}^{N_{\mathrm{T}}L} s_{l}^2 \Big) \\
\mbox{s.t.} &&\hspace*{5mm}\mathbf{\Theta}\in{\cal D}.\notag
\end{eqnarray}
In particular,  $\phi$ acts as  a large penalty factor for penalizing the objective function for any $s_{l}$ that is not equal to $0$ or $1$.
 \end{Thm}

 \emph{\quad Proof: } Please refer to Appendix C for a proof of Theorem \ref{Thm:penalty_method}.

The problem in  (\ref{eqn:equivalent-approx2})  is in the canonical form of difference of convex  (d.c.) functions  programming. Specifically,  $g(s_{l})=\sum_{l=1}^{N_{\mathrm{T}}L} s_{l}^2$ is a concave function and we minimize d.c. functions over a convex constraint set. As a result, we can apply successive convex approximation  \cite{book:SCA_convergence} to obtain a local optimal solution of  (\ref{eqn:equivalent-approx2}).

\begin{table}[t]\vspace*{-16.5mm}\caption{Suboptimal Iterative Resource Allocation Algorithm}\label{table:algorithm2}
\vspace*{-8mm}
\renewcommand\thealgorithm{}
\begin{algorithm} [H]                    
\caption{Successive Convex Approximation}          
\label{alg1}                           
\begin{algorithmic} [1]
\STATE Initialize the maximum number of iterations $L_{\max}$, penalty factor $\phi\gg 0$, iteration index $i=0$, and $s_{l}^{(i)}$
\REPEAT [Loop]
\STATE Solve  (\ref{eqn:equivalent-approx3}) for a given $s_{l}^{(i+1)}$   and obtain the intermediate resource allocation policy $\{\mathbf{W}_k',s_l',P^{\mathrm{U}'}_j,\widetilde{\mathbf{W}}_{k}^{m,n'},\tilde{P}^{\mathrm{U}'}_{j,m,n},{q}_{m,n}'\}$
\STATE Set $s_{l}^{(i+1)}=s_{l}' $ and $ i=i+1$
\UNTIL{ Convergence or $i=L_{\max}$}

\end{algorithmic}
\end{algorithm}
\vspace*{-1.5cm}
\end{table}

\subsubsection{Suboptimal iterative algorithm}\label{subsect:iterative}
 Since  $g(s_{l})$ is a differentiable convex function, inequality
\begin{eqnarray}\label{eqn:global_underestimator}
g(s_{l})&\ge& g(s_{l}^{(i)}) +\nabla_{s_{l}}g(s_{l}^{(i)}) (s_{l}-s_{l}^{(i)}),\forall l\in\{1,\ldots,N_{\mathrm{T}}L\},
\end{eqnarray}
always holds for any feasible point $s_{l}^{(i)}$, where the right hand side of (\ref{eqn:global_underestimator}) is an affine function  \cite{book:convex} and represents a global underestimator  of $g(s_{l})$.

\begin{figure}[t]
 \centering\vspace*{-18mm}
 \begin{minipage}[b]{0.45\linewidth} \hspace*{-1cm}
\includegraphics[width=3.7 in]{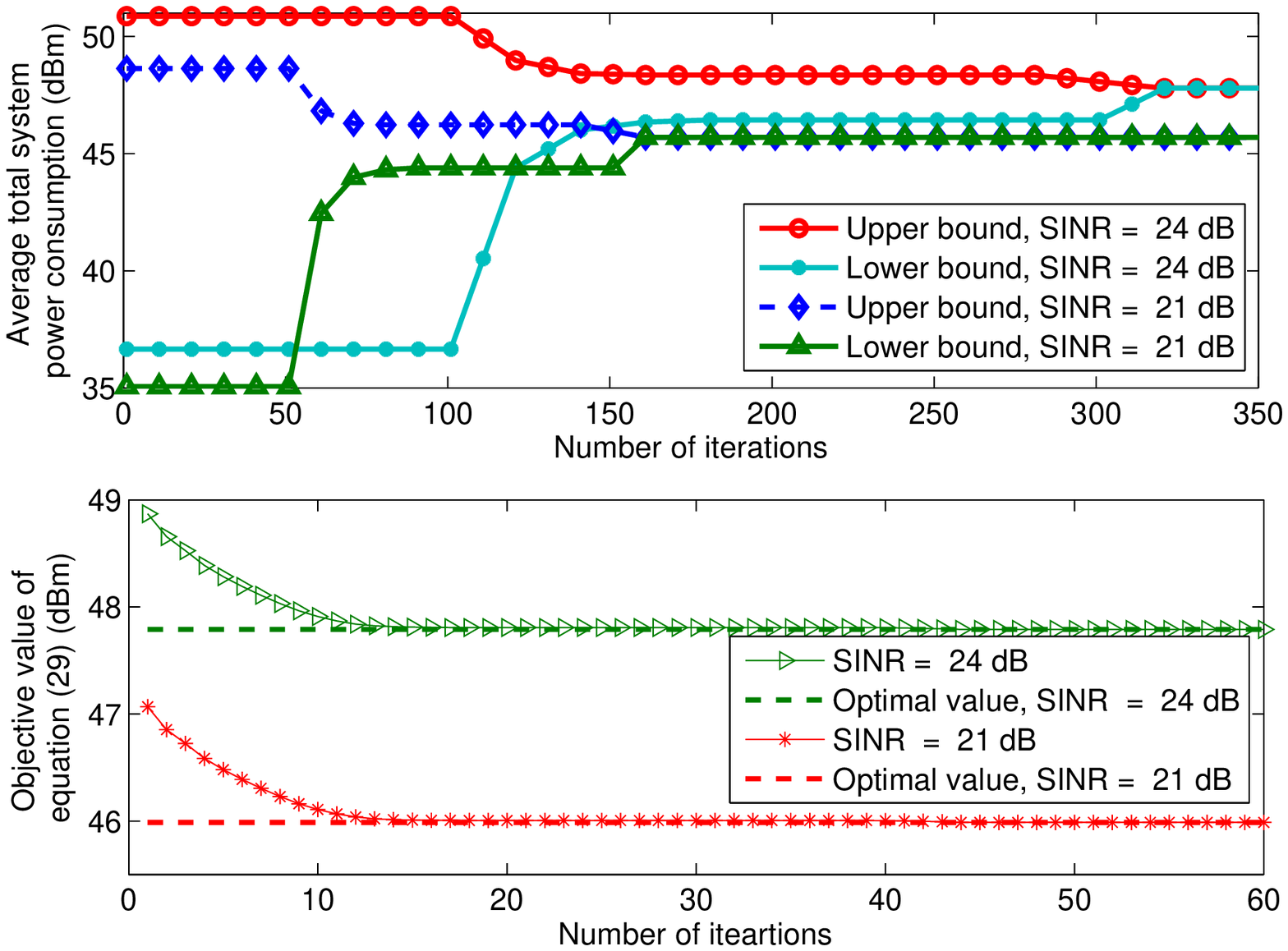}\vspace*{-8mm}
\caption{ Convergence of the proposed iterative algorithms.  } \label{fig:convergence}
 \end{minipage}\hspace*{1.1cm}
 \begin{minipage}[b]{0.45\linewidth} \hspace*{-1cm}
\includegraphics[width=3.7 in]{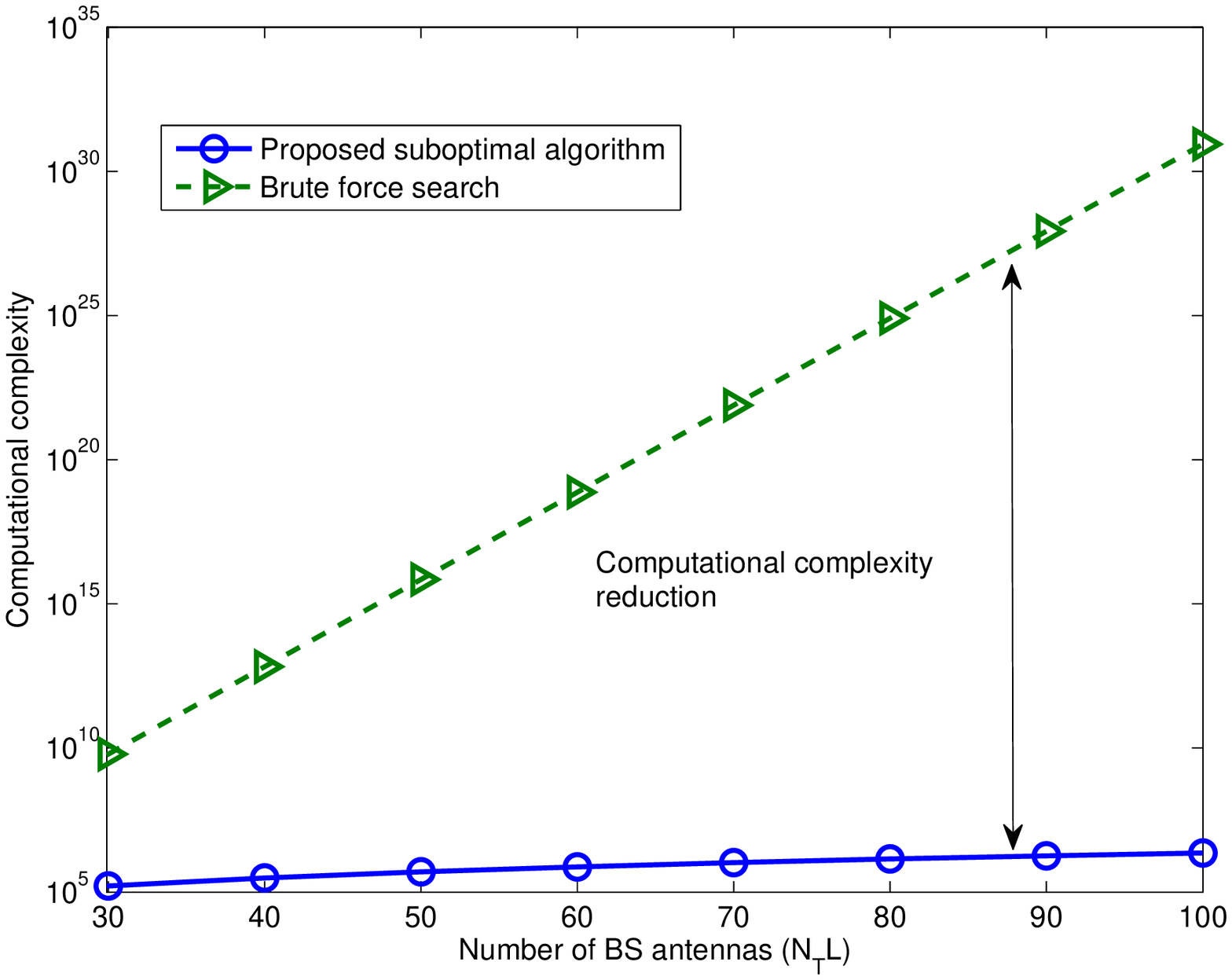}\vspace*{-8mm}
\caption{Computational complexity versus the total number of transmit antennas in the system, $N_{\mathrm{T}}L$.} \label{fig:complexity}
 \end{minipage}\vspace*{-10mm}
\end{figure}

As a result, for any given value of $s_{l}^{(i)}$,  we solve the following optimization problem,
 \begin{eqnarray}\label{eqn:equivalent-approx3}\notag
&&\hspace*{-15mm}   \underset{\mathbf{W}_k\in \mathbb{H}^{N_{\mathrm{T}}L},\widetilde{\mathbf{W}}_{k,b}^l,s_l,P^{\mathrm{U}}_j,\tilde{P}^{\mathrm{U}}_{j,m,n},q_{m,n}}{\mino}\,\, {\cal U}_{\mathrm{TP}}\Big(\hspace*{-0.5mm}\mathbf{W}_k,s_l,P^{\mathrm{U}}_j\hspace*{-0.5mm}\Big)+\phi\Big(\sum_{l=1}^{N_{\mathrm{T}}L}   s_{l} \hspace*{-0.5mm}-\hspace*{-0.5mm}\sum_{l=1}^{N_{\mathrm{T}}L}   (s_{l}^{(i)})^2 \hspace*{-0.5mm}-\hspace*{-0.5mm}2\sum_{l=1}^{N_{\mathrm{T}}L}  s_{l}^{(i)} (s_{l} \hspace*{-0.5mm}-\hspace*{-0.5mm}s_{l}^{(i)})\Big)\\
\mbox{s.t.} &&\hspace*{35mm}\mathbf{\Theta}\in{\cal D},
\end{eqnarray}
which leads to an upper bound of (\ref{eqn:equivalent-approx2}). Then, to tighten the obtained upper bound, we employ an iterative algorithm which is summarized in Table \ref{table:algorithm2}.  First, we initialize the value of $s_{l}^{(i)}$  for iteration index  $i=0$. Then, in each iteration,  we solve (\ref{eqn:equivalent-approx3}) for  given values of $s_{l}^{(i)}$, cf. line 3, and update $s_{l}^{(i+1)}$  with the intermediate solution ${s}_{l}'$, cf. line 4.  The proposed iterative method generates a sequence of feasible solutions $s_{l}^{(i+1)}$ with respect to (\ref{eqn:equivalent-approx2}) by solving the convex upper bound
problem (\ref{eqn:equivalent-approx3}) successively.  As shown in  \cite{book:SCA_convergence}, the proposed suboptimal iterative algorithm converges to a local optimal solution\footnote{By following a similar approach as in the proof of Theorem 1, it can be shown that $\Rank(\mathbf{W}_k)=1$ holds  despite the adopted SDP relaxation.} of (\ref{eqn:equivalent-approx2}) with polynomial time computational complexity.
In fact, the proposed suboptimal algorithm benefits from the convexity of  (\ref{eqn:equivalent-approx3}) and   different numerical methods can be used to efficiently solve  (\ref{eqn:equivalent-approx3}). In particular,  when the \emph{primal-dual path-following interior-point method} is used with a proper choice of kernel(/barrier) function, cf. \cite{JR:complexity1,JR:complexity2},
 the computational complexity of the proposed suboptimal algorithm is $\bigo(L_{\max} (N_{\mathrm{T}}L)^2\ln((N_{\mathrm{T}}L)^2/\epsilon))$ with respect to $N_{\mathrm{T}}L $ for a given solution accuracy $\epsilon>0$
\cite{book:interior_point_complexity}, where $\bigo(\cdot)$ stands for the big-O notation. The computational complexity  is significantly reduced compared to the computational complexity of an exhaustive search which is given by  $\bigo(2^{N_{\mathrm{T}}L}(N_{\mathrm{T}}L\ln(N_{\mathrm{T}}L/\epsilon))$ with respect to $N_{\mathrm{T}}L$, i.e., cf. Figure \ref{fig:complexity}.

\begin{Remark}
The proposed algorithm requires  $s_{l}^{(i)}$ to be a feasible point  for the initialization, i.e., for $i=0$. This point can be easily obtained since the constraints in (\ref{eqn:equivalent-approx2}) span a convex set.
\end{Remark}

\vspace*{-4mm}

\section{Simulation Results}\label{sect:result-discussion}
In this section, we evaluate  the system performance of the proposed resource allocation designs via simulations.
 There are $L=3$ FD radio BSs in the system, which are placed at the corner points of an  equilateral triangle. The inter-site distance between any two FD radio BSs is $250$ meters. The uplink and downlink users are uniformly distributed inside a disc with  radius  $500$ meters centered at the centroid of the triangle.\begin{table}[t]\vspace*{-15mm}\caption{System parameters\vspace*{-5mm}}\label{tab:feedback} \centering
\begin{tabular}{|L|l|}\hline
\hspace*{-1mm}Carrier center frequency and path loss exponent & $1.9$ GHz and  $3.6$ \\
\hline
\hspace*{-1mm}Multipath fading distribution and total noise variance, $\sigma_{\mathrm{z}}^2$ & \mbox{Rayleigh fading} and  $-62$ dBm   \\

\hline

\hspace*{-1mm}Minimum required SINR for uplink user $j$, $\Gamma_{\mathrm{req}_j}^{\mathrm{UL}}$ &  $10$ \mbox{dB}  \\
\hline
\hspace*{-1mm}Power amplifier power efficiency and antenna power consumption in idle mode,  $P^{\mathrm{Idle}}$  & $1/{\varepsilon_{\mathrm{D}}}=1/{\varepsilon_{\mathrm{U}}}=0.2$  and  $0$ dBm   \\
\hline
\hspace*{-1mm}Max. transmit power for downlink and uplink,  $P_{\max_l}^{\mathrm{DL}}$ and $P_{\max_j}^{\mathrm{U}}$ & $48$ dBm and $23$ dBm \\
\hline

\end{tabular}\vspace*{-10mm}
\end{table} We set the  constant weights for the downlink and uplink power consumption as
$  \eta=\zeta_j=1,\forall j\in\{1,\ldots,K_{\mathrm{U}}\}$. The penalty term $\phi$ for the proposed suboptimal algorithm is set to $10 P_{\max_l}^{\mathrm{DL}}$. Also,  $P_0=0$ is adopted in all simulation results\footnote{We note that the value of $P_0$ does not affect the resource allocation algorithm design. }. Unless specified otherwise, we assume $50$ dB of self-interference cancellation\footnote{We assume a balun analog circuit is implemented in the FD radio BSs which can cancel $50$ dB of self-interference \cite{CN:Full_duplex_radio1}. The residual self-interference is handled by the beamforming matrix $\mathbf{W}_k$ via the proposed optimization framework.  } at the FD radio BSs and the circuit power consumption per antenna is $P^{\mathrm{Active}} =30$ dBm. The antenna gains for the BSs and the users are $10$ dBi and $0$ dBi, respectively, and there are $N_{\mathrm{T}}=20$ antennas equipped in each FD BS resulting in $N_{\mathrm{T}}L=60$ antennas in the network. Furthermore, all downlink users require identical minimum SINRs, i.e., $\Gamma_{\mathrm{req}_k}^{\mathrm{DL}}=\Gamma_{\mathrm{req}}^{\mathrm{DL}},\forall k$.   The performance of the proposed algorithms is compared with the performances of the following four baseline systems designed for peak system load when all the available antennas are activated. In particular, we minimize the total system power consumption of all four baseline systems using a similar approach as for the schemes proposed in this paper but set $s_l=1,\forall l\in\{1,\ldots,N_{\mathrm{T}}L\}$. The baseline systems are configured as follows.   \emph{Baseline $1$}: a FD distributed antenna system (FD-DAS); \emph{Baseline $2$}: a HD distributed antenna system (HD-DAS); \emph{Baseline $3$}:  a FD system with co-located antennas (FD-CAS); \emph{Baseline $4$}: a HD  system with co-located antennas (HD-CAS). For the HD communication systems, we adopt  static time division duplex such that   uplink and downlink communication occur in non-overlapping  equal-length time intervals. In other words,  both self-interference and the uplink-to-downlink co-channel interference are avoided. For a fair performance comparison between HD and FD systems, we set $\log_2(1+\Gamma_{\mathrm{req}_j}^{\mathrm{UL}})=1/2\log_2(1+\Gamma_{\mathrm{req}_j}^{\mathrm{UL-HD}})$ and $\log_2(1+\Gamma_{\mathrm{req}_j}^{\mathrm{DL}})=1/2\log_2(1+\Gamma_{\mathrm{req}_j}^{\mathrm{DL-HD}})$ such that  the minimum required SINRs for the uplink users, $\Gamma_{\mathrm{req}_j}^{\mathrm{UL-HD}}$, and downlink users, $\Gamma_{\mathrm{req}_j}^{\mathrm{DL-HD}}$, become $\Gamma_{\mathrm{req}_j}^{\mathrm{UL-HD}} =(1+\Gamma_{\mathrm{req}_j}^{\mathrm{UL}})^2-1 $ and $\Gamma_{\mathrm{req}_j}^{\mathrm{DL-HD}} =(1+\Gamma_{\mathrm{req}}^{\mathrm{DL}})^2-1 $, respectively, to account for the penalty due to the loss in spectral efficiency of the HD protocol. Also,  the  power consumption of downlink and uplink transmission in the objective function of the HD systems is reduced by a factor of two as at a given time either uplink or downlink transmission is performed. For the  CAS, we assume that there is only one BS located  at the center of the system, which is equipped with the same number of antennas as all FD BSs in the distributed setting combined, i.e., $N_{\mathrm{T}}L$. Furthermore, for all baseline systems,  we remove the maximum transmit power constraints imposed for the downlink and uplink transmissions, i.e., constraints C3 and C4. The key parameters adopted in the simulations are provided in Table \ref{tab:feedback}.

\vspace*{-6mm}

\subsection{Convergence  and Computational Complexity of the Proposed Iterative Algorithms}
Figure \ref{fig:convergence} illustrates the convergence of
the proposed optimal and suboptimal algorithms  for  different minimum required SINRs for downlink users, $\Gamma_{\mathrm{req}}^{\mathrm{DL}}$.  There are $K_{\mathrm{D}}=4$ downlink users and $K_{\mathrm{U}}=2$ uplink users in the system. It can be seen from the upper half of Figure \ref{fig:convergence} that
the proposed optimal algorithm in Table I converges to the optimal solution  in less than $350$ iterations, i.e.,  the upper bound value meets the lower bound value. On the other hand, from the lower half of Figure \ref{fig:convergence}, we observe that the suboptimal algorithm converges to a local optimal value after less than $20$ iterations. In the sequel, we show the performance of the suboptimal iterative algorithm for $20$ iterations.

 Figure \ref{fig:complexity} compares the computational complexity of the brute force approach with that of the proposed suboptimal algorithm\footnote{ The proposed optimal algorithm may have the same computational complexity as the brute force approach  in the worst case scenario although this seldom happens in practice.   } for $20$ iterations and solution accuracy $\epsilon=0.1$. The system setting is identical to the scenario in Figure \ref{fig:convergence} and the results are computed based on the big-O complexity analysis in Section \Rmnum{4}. As can be observed, the proposed suboptimal resource allocation algorithm requires a significantly lower   computational complexity compared to the brute force approach, especially for large numbers of antennas.

\begin{figure}[t]\label{fig:p_SINR}
 \centering\vspace*{-18mm}
\includegraphics[width=4.4 in]{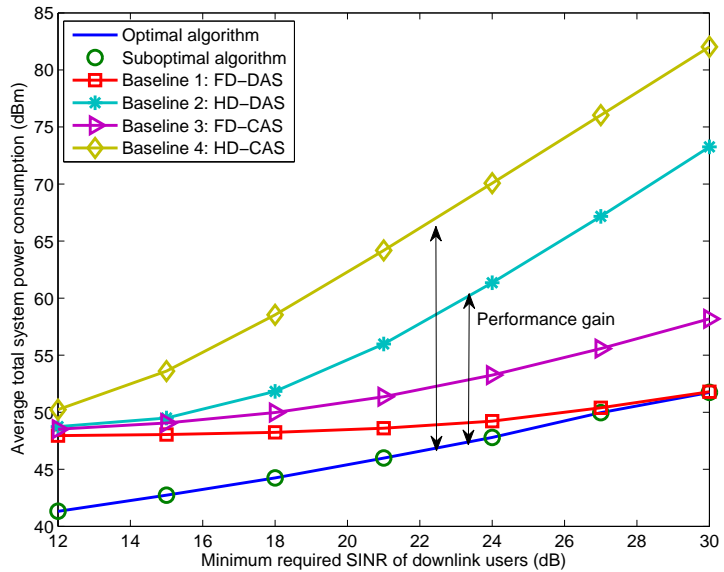}\vspace*{-6mm}
\caption{Average total system power consumption  (dBm) versus the  minimum required SINRs for the downlink users, $\Gamma_{\mathrm{req}}^{\mathrm{DL}}$, for different  systems. The double-sided arrows indicate the performance gain of the proposed FD system compared to traditional HD communication systems.}\vspace*{-8mm}\label{fig:p_SINR}
\end{figure}
\vspace*{-4mm}
\subsection{Average Total System Power Consumption}
\begin{figure}[t]
 \centering\vspace*{-18mm}
 \begin{minipage}[b]{0.45\linewidth} \hspace*{-1cm}
\includegraphics[width=3.7 in]{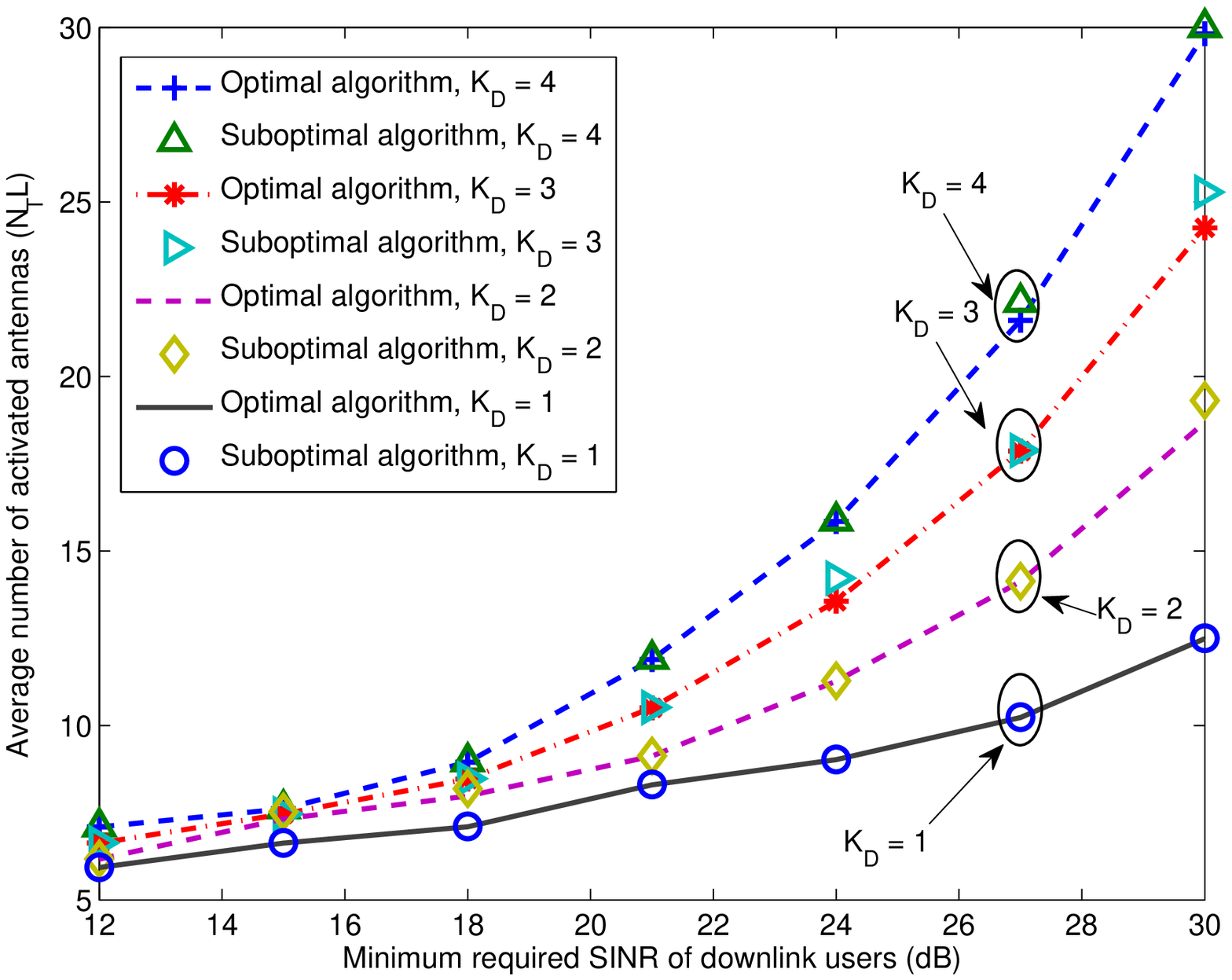}\vspace*{-8mm}
\caption{Average number of activated antennas versus $\Gamma_{\mathrm{req}}$.  }\label{fig:nt_SINR}
 \end{minipage}\hspace*{1.1cm}
 \begin{minipage}[b]{0.45\linewidth} \hspace*{-1cm}
\includegraphics[width=3.7 in]{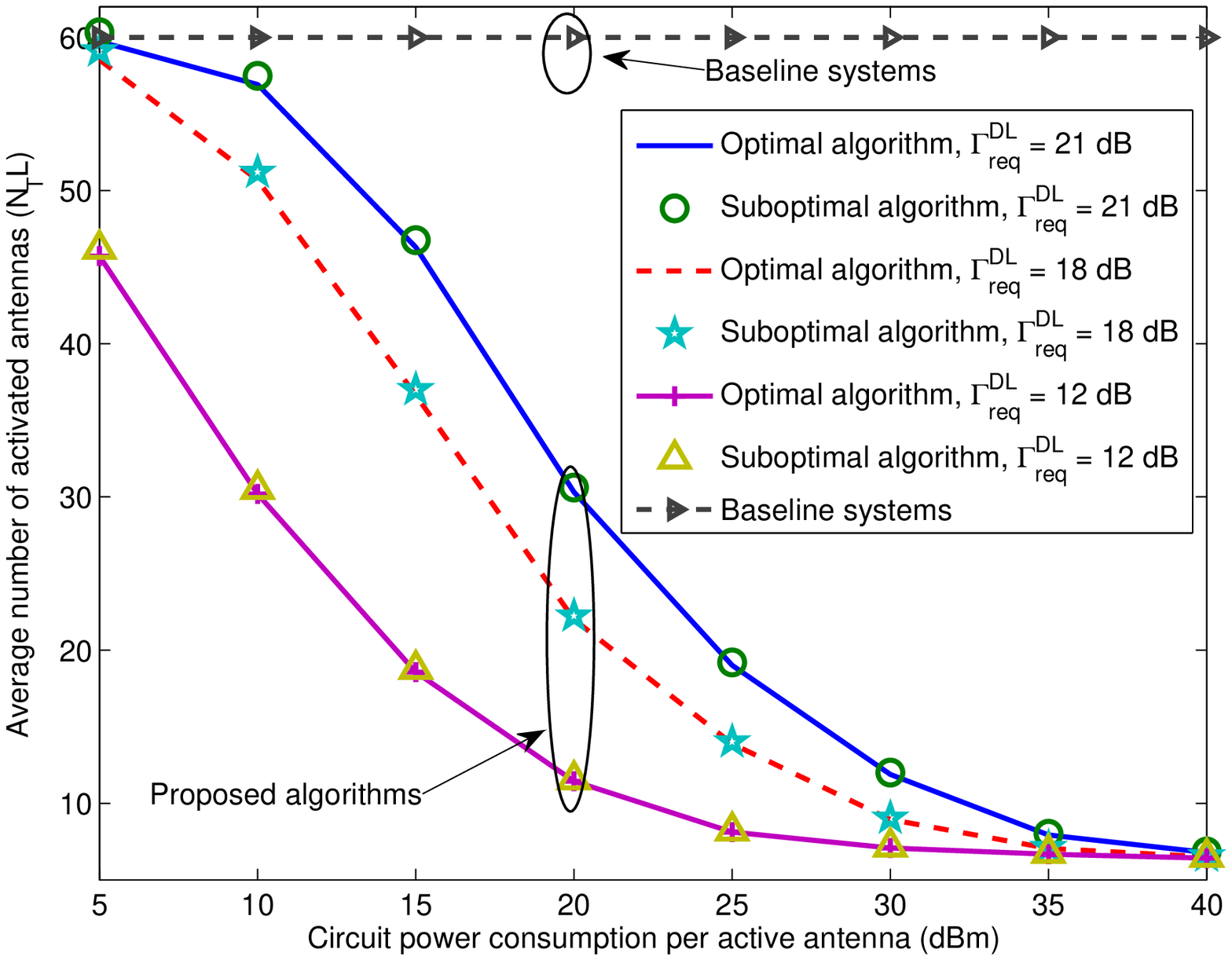}\vspace*{-8mm}
\caption{Average number of activated antennas versus circuit power consumption per active antenna. } \label{fig:nt_pc}
 \end{minipage}\vspace*{-10mm}
\end{figure}

In Figure \ref{fig:p_SINR}, we study the average total system power consumption versus the minimum required SINRs of the downlink users, $\Gamma_{\mathrm{req}}^{\mathrm{DL}}$.  There are $K_{\mathrm{D}}=4$ downlink users and $K_{\mathrm{U}}=2$ uplink users in the system. It can be observed  that the average total system power consumption increases gradually with $\Gamma_{\mathrm{req}}^{\mathrm{DL}}$. In fact, as the QoS requirements of the downlink users become more stringent,  a higher downlink  transmit power is needed to fulfill the requirement. At the same time, the self-interference power increases with the downlink transmit power. Thus, the FD radio BSs have to utilize more degrees of freedom for self-interference suppression, and as a consequence,  less degrees of freedom are available for reducing the total system power consumption.  On the other hand, the proposed suboptimal iterative resource allocation algorithm offers practically the same performance as the optimal algorithm for the considered scenario. As can be observed, the two proposed algorithms facilitate significant power savings compared to all baseline system architectures (which activate always all available antennas),  especially for low to moderate system loads, i.e., $\Gamma_{\mathrm{req}}^{\mathrm{DL}}\le21$ dB. Indeed,  activating all antennas may not be beneficial for the total system power consumption when the load of the system is relatively small, since in this case, the power consumption caused by an extra antenna circuit outweighs the power reduction for information  transmission offered by the extra activated antenna. Nevertheless,  the performance gap between the two proposed algorithms and baseline system $1$  diminishes as  the minimum required SINRs for the downlink users increase. In particular, the BSs are forced to transmit with high power to satisfy the more stringent QoS requirements when the number of activated antennas is small. As a result, the two proposed algorithms have to activate more antennas, cf. also Figure \ref{fig:nt_SINR}, for improving the power efficiency of the system which yields  a similar resource allocation as baseline system $1$.  Additionally,  the two proposed  algorithms outperform HD baseline systems $2$ and $4$ by a considerable margin. As can be seen, in the HD systems, an exceedingly large system power consumption is required to meet the more   stringent minimum required downlink SINRs to compensate for the spectral efficiency loss inherent to the HD protocol. Furthermore,  the distributed antennas deployed in the proposed systems provide spatial diversity across the network  which shortens the distance between transmitters and receivers. This  accounts for the power saving  enabled by the two proposed algorithms compared to  baseline CASs $3$ and $4$.

\begin{figure}[t]
 \centering\vspace*{-18mm}
\includegraphics[width=4.4 in]{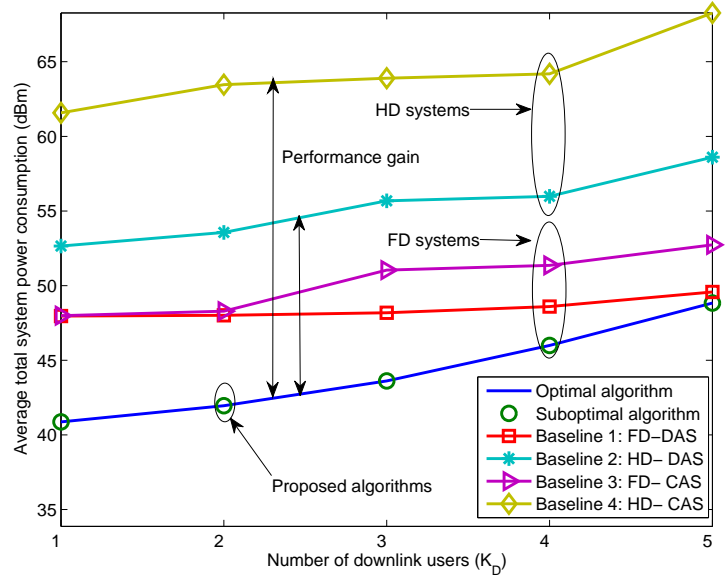}\vspace*{-6mm}
\caption{Average total system power consumption  (dBm) versus the  number of  downlink users for $\Gamma_{\mathrm{req}}^{\mathrm{DL}}=21$ dB and  different  systems. The double-sided arrows indicate the performance gain of the proposed FD protocol compared to the HD protocol.}\vspace*{-8mm}\label{fig:p_users}
\end{figure}

Figure \ref{fig:p_users} depicts the average total system power consumption versus the number of downlink users for a minimum required downlink SINR of $\Gamma_{\mathrm{req}}^{\mathrm{DL}}=21$ dB.   There are  $K_{\mathrm{U}}=2$ uplink users in the system. It is observed that the average total system power consumption increases with the number of downlink users.  As more downlink users request communication services from the system, more QoS constraints are imposed on the optimization problem in  (\ref{eqn:cross-layer}) which reduces the size of the feasible solution set and thus results in a higher total system power consumption. In addition, the  two proposed resource allocation algorithms outperform all baseline schemes due to the adopted optimization framework and the  distributed antenna architecture.

\vspace*{-6mm}
\subsection{Average Number of Activated Antennas}
In Figure \ref{fig:nt_SINR}, we study the average number of  activated antennas  versus the minimum required downlink SINR,  $\Gamma_{\mathrm{req}}^{\mathrm{DL}}$, for different  numbers of downlink users.  It can be observed that the average number of activated antennas   increases with increasing  minimum required SINR for the downlink users. Although activating an extra antenna for signal transmission and reception consumes extra power in the circuit, i.e., $P^{\mathrm{Active}}- P^{\mathrm{Idle}}>0$, a larger number of activated antennas increases the degrees of freedom of the system which is beneficial if the QoS constraints are stringent. Specifically, with more antennas,  the direction of beamforming
matrix $\mathbf{W}_k$ can be   more accurately steered towards downlink user $k$  which substantially reduces the necessary downlink transmit power to achieve a certain QoS.  Moreover, the reduced downlink transmit power also decreases the self-interference which in turn reduces the required uplink transmit power. In fact, for a small number of activated antennas, the FD radio BSs are required to transmit with exceedingly high power if $\Gamma_{\mathrm{req}}^{\mathrm{DL}}$ is large. As a result, the FD radio BSs prefer to activate more antennas to improve the power efficiency of information transmission, when the cost of activating extra antennas is less than the associated potential transmit power saving. On the other hand, it can be observed that the proposed schemes activate more antennas when more downlink users are in the system. In fact, the downlink co-channel interference increases with  the number of downlink users. Furthermore, the co-channel interference cannot be suppressed by   simply increasing the downlink transmit power for all downlink users. Thus, extra spatial degrees of freedom are beneficial  for decreasing the system power consumption.

In Figure \ref{fig:nt_pc}, we show  the average number of  activated antennas    versus the circuit power consumption per active antenna, $P^{\mathrm{Active}}$ (dBm), for different    minimum required SINRs for the downlink users.  It is expected that the FD radio BSs prefer to activate more antennas when the circuit power consumption per antenna is small or the SINR requirements of the downlink users are demanding, since in this case,  the power savings achieved  by activating extra antennas surpasses the corresponding circuit power consumption.  On the contrary, when the circuit power consumption per antenna is high, the FD radio BSs become more conservative in activating antennas since using a large number of antennas may no longer be beneficial to the overall system power consumption.
\vspace*{-4mm}
\section{Conclusions}\label{sect:conclusion}
In this paper, we formulated the resource allocation algorithm design for power efficient distributed  FD  antenna networks as a mixed combinatorial and non-convex optimization problem, where the antenna circuit power consumption and the QoS requirements of the uplink and downlink users were taken into account. Applying the  generalized Bender's decomposition, we developed an optimal  iterative resource allocation algorithm for solving the problem optimally. In addition, a polynomial time computational complexity suboptimal algorithm  was also proposed to strike a balance between computational complexity and optimality. Simulation results showed that the proposed suboptimal iterative resource allocation algorithm   approaches the optimal performance in a small number of iterations. Furthermore, our results unveiled the substantial power savings enabled in FD radio distributed antennas networks by dynamically switching off a subset of the available antennas; an exceedingly large number of activated antennas
may not be a cost effective solution for reducing the total system power consumption when the QoS requirements of the users are not stringent.

\section*{Appendix}
\subsection{Proof of Theorem \ref{thm:rankone_condition}}
\label{appendix:proof_thm}
We start the proof by rewriting the Lagrangian function of the primal problem in (\ref{eqn:primal_problem}) in terms of the beamforming matrix $\mathbf{W}_k$:
\begin{eqnarray}
{\cal L}\Big(\mathbf{\Theta},\mathbf{\Phi}\Big)&=&\sum_{k=1}^{K_{\mathrm{D}}}\Tr(\mathbf{A}_k\mathbf{W}_k)-
\sum_{k=1}^{K_{\mathrm{D}}}\Tr\Big(
\big(\mathbf{Z}_k+\frac{\alpha_k\mathbf{H}_{\mathrm{D}_k}}{\Gamma_{\mathrm{req}_k}^{\mathrm{DL}}}\big)\mathbf{W}_k\Big)+\Delta \\
\label{eqn:A_k}
\mbox{and}\quad\mathbf{A}_k&=&\eta\varepsilon_\mathrm{D}\mathbf{I}_{N_{\mathrm{T}L}}+\sum_{j\neq k}^{K_{\mathrm{D}}}\alpha_j\mathbf{H}_{\mathrm{D}_j}+\sum_{l=1}^{N_{\mathrm{T}}L}\rho\mathbf{R}_l+\sum_{m=1}^{N_{\mathrm{T}}L} \sum_{n=1}^{N_{\mathrm{T}}L}\Big(\mathbf{D}_{\mathrm{C}_{17_{k,m,n}}}-\mathbf{D}_{\mathrm{C}_{16_{k,m,n}}}\Big).
\end{eqnarray}
$\Delta$ denotes the collection of  variables that are independent of $\mathbf{W}_k$. For convenience,  the optimal primal and  dual variables of the SDP relaxed version of  (\ref{eqn:primal_problem}) are denoted by the corresponding variables with an asterisk  superscript. By exploiting the Karush-Kuhn-Tucker (KKT) optimality conditions, we obtain the following  equations:
\begin{eqnarray}\vspace*{-2mm}
\hspace*{-3mm}\mathbf{Z}_k^*\hspace*{-3mm}&\succeq&\hspace*{-3mm}\mathbf{0},\,\,\alpha_k^*\ge 0,\,\forall k, \label{eqn:dual_variables}\\
\hspace*{-3mm}\mathbf{Z}_k^*\mathbf{W}_k^*\hspace*{-3mm}&=&\hspace*{-3mm}\mathbf{0},\label{eqn:KKT-complementarity}\\
\hspace*{-3mm}\mathbf{Z}_k^*\hspace*{-3mm}&=&\hspace*{-3mm}\mathbf{A}_{k}^*-\frac{\alpha_k^*\mathbf{H}_{\mathrm{D}_k}}
{\Gamma_{\mathrm{req}_k}^{\mathrm{DL}}},
\label{eqn:lagrangian_gradient}
\end{eqnarray}
where $\mathbf{A}_{k}^*$ in (\ref{eqn:lagrangian_gradient}) is obtained by substituting the optimal dual variables $\boldsymbol \Phi^*$ into (\ref{eqn:A_k}). From (\ref{eqn:KKT-complementarity}), we know that the optimal beamforming matrix $\mathbf{W}^*_k$ is a rank-one matrix when  $\Rank(\mathbf{Z}^*_k)=N_{\mathrm{T}}L-1$. In particular, $\mathbf{W}^*_k$ is required to lie in the null space spanned by $\mathbf{Z}^*_k$ for $\mathbf{W}^*_k\ne\zero$. As a result, by revealing the structure of $\mathbf{Z}^*_k$, we can study the rank of beamforming matrix $\mathbf{W}^*_k$. In the following, we first show by contradiction that $\mathbf{A}_k^*$ is a positive definite matrix with probability one. To this end, we focus on the dual problem in (\ref{eqn:dual}). For a given set of optimal dual variables, $\boldsymbol \Phi^*$,  and a subset of optimal primal variables, $\{s_l^*,P^{\mathrm{U}*}_j,\widetilde{\mathbf{W}}_{k}^{m,n*},\tilde{P}^{\mathrm{U}*}_{j,m,n},{q}_{m,n}^*\}$,  the dual problem in (\ref{eqn:dual}) can be written as\vspace*{-2mm}
\begin{eqnarray}\hspace*{-2mm}\label{eqn:dual2}\vspace*{-2mm}
\,\,\underset{\mathbf{W}_k\in\mathbb{H}^{N_{\mathrm{T}}}}{\mino} \,\, {\cal L}\Big(\hspace*{-0.5mm}\mathbf{\Theta},\mathbf{\Phi}^*\hspace*{-0.5mm}\Big).
\end{eqnarray}
Suppose $\mathbf{A}_k^*$ is negative semi-definite, i.e., $\mathbf{A}_k^*\preceq\zero$,  then we can construct a beamforming matrix $\mathbf{W}_k=r\mathbf{\tilde w}_k\mathbf{\tilde w}_k^H$ as one of the solutions of (\ref{eqn:dual2}), where $r>0$ is a scaling parameter and $\mathbf{\tilde w}_k$ is the eigenvector corresponding to one of the non-positive eigenvalues of $\mathbf{A}_k^*$. We substitute $\mathbf{W}_k=r\mathbf{\tilde w}_k\mathbf{\tilde w}_k^H$ into (\ref{eqn:dual2}) which yields
\begin{equation}\vspace*{-2mm}
{\cal L}\Big(\mathbf{\Theta},\mathbf{\Phi}\Big)=\underbrace{\sum_{k=1}^{K_{\mathrm{D}}}\Tr(r\mathbf{A}_k^*\mathbf{\tilde w}_k\mathbf{\tilde w}_k^H)}_{\le 0}-r\sum_{k=1}^{K_{\mathrm{D}}}\Tr\Big(\mathbf{\tilde w}_k\mathbf{\tilde w}_k^H
\big(\mathbf{Z}_k^*+\frac{\alpha_k^*\mathbf{H}_{\mathrm{D}_k}}{\Gamma_{\mathrm{req}_k}^{\mathrm{DL}}}\big)\Big)+\Delta.
\end{equation}
 Besides, constraint C1 is satisfied with equality for the optimal solution and thus $\alpha_k>0$. Furthermore,  since the channel vectors of the downlink users, i.e.,   $\mathbf{h}_{\mathrm{D}_k}$, $\forall k\in\{1,\ldots,K_{\mathrm{D}}\}$, are assumed to be statistically independent, we obtain $-r\sum_{k=1}^{K_{\mathrm{D}}}\Tr\Big(\mathbf{\tilde w}_k\mathbf{\tilde w}_k^H
\big(\mathbf{Z}_k^*+\frac{\alpha_k^*\mathbf{H}_{\mathrm{D}_k}}{\Gamma_{\mathrm{req}_k}^{\mathrm{DL}}}\big)\Big)\rightarrow -\infty$ when we set $r\rightarrow \infty$. Thus, the dual optimal value  becomes unbounded from below. Yet, the optimal value of the primal problem in (\ref{eqn:primal_problem}) is non-negative for $\Gamma_{\mathrm{req}_k}^{\mathrm{DL}}>0$ which leads to a contradiction as strong duality does not hold. Therefore, for the optimal solution,  $\mathbf{A}_k^*$ is a positive definite matrix with probability one and $\Rank(\mathbf{A}_k^*)=N_{\mathrm{T}}L$, i.e., $\mathbf{A}_k^*$ has full rank.

Then, by exploiting (\ref{eqn:lagrangian_gradient}) and basic rank inequality results, we have the following implication:
\begin{eqnarray}\notag
\hspace*{-3mm}&&\hspace*{-2mm}\Rank(\mathbf{Z}^*_k)+\Rank\big(\alpha_k^*\frac{\mathbf{H}_{\mathrm{D}_k}}{\Gamma_{\mathrm{req}_k}^{\mathrm{DL}}}\big)
\ge\Rank\big(\mathbf{Z}^*_k+\alpha_k^*\frac{\mathbf{H}_{\mathrm{D}_k}}{\Gamma_{\mathrm{req}_k}^{\mathrm{DL}}}\big)=\Rank(\mathbf{A}_k^*)=N_\mathrm{T}L\\
\hspace*{-3mm}&\Rightarrow &\hspace*{-2mm}
 \Rank(\mathbf{Z}^*_k)\ge N_{\mathrm{T}}L-1.
\end{eqnarray}
 Furthermore, $\mathbf{W}_k^*\ne\mathbf{0}$ is required to satisfy  C1 for $\Gamma_{\mathrm{req}_k}^{\mathrm{DL}}>0$. Thus, $\Rank(\mathbf{Z}^*_k)=N_{\mathrm{T}}L-1$ and $\Rank(\mathbf{W}^*_k)=1$ hold with probability one.  \qed
\vspace*{-3mm}
\subsection{Proof of Proposition \ref{thm:supporting_plane}}
We start the proof by studying the solution of the SDP relaxed version of  (\ref{eqn:primal_problem}) via its dual problem in (\ref{eqn:dual}). For a given set of optimal dual variables  $\boldsymbol \Phi(i)$, we have $\boldsymbol \Theta(i)$\vspace*{-3mm}
\begin{eqnarray}\notag\label{eqn:proof_propo1}
\notag
\hspace*{-3.5mm}&=&\hspace*{-2.5mm} \arg\underset{\mathbf{\Theta}}{\min}\,\,{\cal U}_{\mathrm{TP}}\Big(\hspace*{-0.5mm}\mathbf{W}_k,s_l,
P^{\mathrm{U}}_j\hspace*{-0.5mm}\Big)\hspace*{-0.5mm}+\hspace*{-0.5mm}
f_1(\mathbf{\Theta},\mathbf{\Phi}(i))\notag\\
\hspace*{-3.5mm}&&\hspace*{-2.5mm}\hspace*{-1.5mm}+ \sum_{k=1}^{K_{\mathrm{D}}}\sum_{l=1}^{N_{\mathrm{T}}L} \rho_{l}\Tr(\mathbf{W}_k\mathbf{R}_l)+ \sum_{m=1}^{N_{\mathrm{T}}L} \sum_{n=1}^{N_{\mathrm{T}}L} \kappa_{m,n} q_{m,n}+ \varphi_{m,n} q_{m,n}- \omega_{m,n} q_{m,n},\label{eqn:supporting_plane_eq1}
\end{eqnarray}
where the first equality is due to the KKT conditions of the SDP relaxed version of  (\ref{eqn:primal_problem}). On the other hand, we can rewrite function $\xi( \boldsymbol{\Phi}(t),s_{l,k}),t\in\{1,\ldots,i\}$, in (\ref{eqn:opt_master_1}) as $\xi( \boldsymbol{\Phi}(t),s_{l,k})$
\begin{eqnarray}\notag\small
\hspace*{-3mm}&=&\hspace*{-3mm} \Bigg\{\underset{\mathbf{\Theta}}{\mino}\,\,{\cal U}_{\mathrm{TP}}\Big(\hspace*{-0.5mm}\mathbf{W}_k,s_l,
P^{\mathrm{U}}_j\hspace*{-0.5mm}\Big)\hspace*{-0.5mm}+\hspace*{-0.5mm}
f_1(\mathbf{\Theta},\mathbf{\Phi}(i))\hspace*{-0.5mm}+\hspace*{-0.5mm} \sum_{k=1}^{K_{\mathrm{D}}}\sum_{l=1}^{N_{\mathrm{T}}L} \rho_{l}\Tr(\mathbf{W}_k\mathbf{R}_l)\hspace*{-0.5mm}+\hspace*{-0.5mm} \sum_{m=1}^{N_{\mathrm{T}}L} \sum_{n=1}^{N_{\mathrm{T}}L} \hspace*{-0.5mm}\kappa_{m,n} q_{m,n}\notag\\
&&+ \varphi_{m,n} q_{m,n}- \omega_{m,n} q_{m,n}\Bigg\}+\sum_{m=1}^{N_{\mathrm{T}}L} \sum_{n=1}^{N_{\mathrm{T}}L} \omega_{m,n} (s_n+s_m-1)  - \kappa_{m,n} s_m - \varphi_{m,n} s_n\notag\\
&&-\sum_{k=1}^{K_{\mathrm{D}}}\sum_{l=1}^{N_{\mathrm{T}}L}  s_l P_{\max_l}^{\mathrm{DL}}
\label{eqn:supporting_plane_eq2}
\end{eqnarray}
The difference between (\ref{eqn:proof_propo1}) and (\ref{eqn:supporting_plane_eq2}) is a constant offset. Thus,  $\boldsymbol \Theta(t)$ is also the solution for the minimization in the master problem in (\ref{eqn:supporting_plane_eq2}) for the $t$-th constraint in (\ref{eqn:supporting_plane_constraint1}). The same approach can be adopted to prove that the solution of (\ref{eqn:FP}) is also the solution of (\ref{eqn:opt_master_2}). \qed
\vspace*{-3mm}
 \subsection{Proof of Theorem \ref{Thm:penalty_method}}
 \label{appendix:thm2}
We start the proof of Theorem \ref{Thm:penalty_method} by using the  \emph{abstract Lagrangian duality} \cite{JR:DC_programming,book:perturbation_function,book:Conjugate_duality}. In particular, the optimization problem  in (\ref{eqn:equivalent-dc-constraint}) can be written as \begin{eqnarray}\label{eqn:abstrct_Lagrangian}&&\hspace*{-4mm}
\underset{{\mathbf\Theta}\in {\cal D}}{\mino}\,\,\,\underset{\phi\ge 0}{\maxo}\quad \overline{\cal L}(\mathbf{\Theta} ,\phi) \end{eqnarray}
where
\begin{eqnarray}\overline{\cal L}(\mathbf{\Theta} ,\phi)={\cal U}_{\mathrm{TP}}\Big(\hspace*{-0.5mm}\mathbf{W}_k,s_l,P^{\mathrm{U}}_j\hspace*{-0.5mm}\Big)+\phi\Big(\sum_{l=1}^{N_{\mathrm{T}}L}   s_{l} -\sum_{l=1}^{N_{\mathrm{T}}L} s_{l}^2 \Big)
 \end{eqnarray}
 and the dual problem of (\ref{eqn:equivalent-dc-constraint}) is given by
 \begin{eqnarray}&&\underset{\phi\ge 0}{\maxo}\quad\underset{{\mathbf\Theta}\in {\cal D}}{\mino}\,\,\,\quad \overline{\cal L}(\mathbf{\Theta} ,\phi).
 \end{eqnarray}
   For notational simplicity, we define
 \begin{eqnarray}
 \Omega(\phi)= \underset{\mathbf{\Theta}\in {\cal D}}{\mino}\quad \overline{\cal L}(\mathbf{\Theta} ,\phi).
 \end{eqnarray}
  Then, we have the following inequalities:
  \begin{subequations}
  \begin{eqnarray}\label{eqn:min_max_bound1}
 \underset{\phi\ge0}{\maxo} \quad\Omega(\phi)&=&  \underset{\phi\ge0}{\maxo} \quad \underset{\mathbf{\Theta}\in {\cal D}}{\mino}\quad \overline{\cal L}(\mathbf{\Theta} ,\phi)\\
 &\stackrel{\mbox{(a)}}{\le }&  \underset{\mathbf{\Theta}\in {\cal D}}{\mino}\quad \underset{\phi\ge0}{\maxo} \quad \overline{\cal L}(\mathbf{\Theta} ,\phi)= \mbox{(\ref{eqn:equivalent-dc-constraint})},
  \end{eqnarray}
  \end{subequations}
  where (a) is due to the weak duality \cite{book:convex}. We note that $\sum_{l=1}^{N_{\mathrm{T}}L}   s_{l} -\sum_{l=1}^{N_{\mathrm{T}}L} s_{l}^2 \ge 0$  for $\mathbf{\Theta}\in{\cal D}$ such that $ \overline{\cal L}(\mathbf{\Theta} ,\phi)$ is a monotonically increasing function in $\phi$. In other words, $\Omega(\phi)$ is increasing in $\phi$ and is bounded from above by the optimal value of (\ref{eqn:abstrct_Lagrangian}).
  Suppose the optimal solution for (\ref{eqn:min_max_bound1}) is denoted as $\phi_0^*$ and $\mathbf{\Theta}^*=\{\mathbf{W}_k,s_l,P^{\mathrm{U}}_j,\widetilde{\mathbf{W}}_{k,b}^l,\tilde{P}^{\mathrm{U}}_{j,m,n},q_{m,n}\}$, where $0\le\phi_0^*\le \infty$. Then, we  study the following two cases for the solution structure of
   (\ref{eqn:min_max_bound1}).  In the first case, we assume $\sum_{l=1}^{N_{\mathrm{T}}L}   s_{l} -\sum_{l=1}^{N_{\mathrm{T}}L} s_{l}^2 = 0$ for (\ref{eqn:min_max_bound1}). As a result, $\mathbf{\Theta}^*$ is also a feasible solution to (\ref{eqn:equivalent-dc-constraint}). Subsequently, we substitute
  $\mathbf{\Theta}^*$ into the optimization problem in  (\ref{eqn:equivalent-dc-constraint}) which yields:
  \begin{eqnarray}\label{eqn:min_max_lower_bound}
  \Omega(\phi_0^*)={\cal U}_{\mathrm{TP}}\Big(\hspace*{-0.5mm}\mathbf{W}_k,s_l,P^{\mathrm{U}}_j\hspace*{-0.5mm}\Big) \ge \mbox{(\ref{eqn:equivalent-dc-constraint})}.
  \end{eqnarray}
  By utilizing (\ref{eqn:min_max_bound1}) and (\ref{eqn:min_max_lower_bound}),  we can conclude that
 \begin{eqnarray}\underset{\mathbf{\Theta}\in {\cal D}}{\mino}\,\,\,\underset{\phi\ge 0}{\maxo}\quad \overline{\cal L}(\mathbf{\Theta} ,\phi) = \underset{\phi\ge 0}{\maxo}\quad \underset{\mathbf{\Theta}\in {\cal D}}{\mino}\quad \overline{\cal L}(\mathbf{\Theta} ,\phi)
 \end{eqnarray}
 must hold for $\sum_{l=1}^{N_{\mathrm{T}}L}   s_{l} -\sum_{l=1}^{N_{\mathrm{T}}L} s_{l}^2 = 0$.  Furthermore, the monotonicity of $\Omega(\phi)$  with respect to  $\phi$ implies that\vspace*{-4mm}
  \begin{eqnarray}\label{eqn}
   \Omega(\phi)= \mbox{(\ref{eqn:equivalent-dc-constraint})},\quad \forall  \phi\ge \phi_0^*,
 \end{eqnarray}
 and the result of Theorem \ref{Thm:penalty_method} follows immediately.

Now, we study the case of
$\sum_{l=1}^{N_{\mathrm{T}}L}   s_{l} -\sum_{l=1}^{N_{\mathrm{T}}L} s_{l}^2 > 0$  at the optimal solution for (\ref{eqn:min_max_bound1}). The optimization problem $ \underset{\phi\ge0}{\maxo} \quad\Omega(\phi)\rightarrow \infty$ is unbounded from above due to the  monotonicity of function $\Omega(\phi)$ with respect to $\phi$. This contradicts the inequality in (\ref{eqn:min_max_bound1}) as (\ref{eqn:equivalent-dc-constraint}) is finite and positive. Thus, for the optimal solution,  $\sum_{l=1}^{N_{\mathrm{T}}L}   s_{l} -\sum_{l=1}^{N_{\mathrm{T}}L} s_{l}^2 = 0$ holds and the result of Theorem \ref{Thm:penalty_method} follows immediately from the first considered case.  \qed
\vspace*{-4mm}
\bibliographystyle{IEEEtran}

\end{document}